\documentclass[prb,twocolumn,superscriptaddress,showpacs]{revtex4}
\usepackage{amsmath,amssymb,graphicx,epsfig,color, float}
\usepackage{epstopdf}

\begin{document}
\title{Spin Response to Localized Pumps: Exciton Polaritons Versus Electrons and Holes}

\author{Vincent Sacksteder IV}
\email{vincent@sacksteder.com}
\affiliation{Division of Physics, Royal Holloway University of London}

\author{A. A. Pervishko}
\affiliation{Division of Physics and Applied Physics, Nanyang Technological University 637371, Singapore}

\author{I. A. Shelykh}
\affiliation{Division of Physics and Applied Physics, Nanyang Technological University 637371, Singapore}
\affiliation{Science Institute, University of Iceland, Dunhagi-3, IS-107, Reykjavik, Iceland}
\affiliation{National Research University for Information Technology, Mechanics and Optics (ITMO), St. Petersburg 197101, Russia}

\date{\today}

\begin{abstract}
Polariton polarization  can be described in terms of a pseudospin which can be oriented along the $x,\,y,$ or $z$ axis, similarly to electron and hole spin.   Unlike electrons and holes where time-reversal symmetry requires that the spin-orbit interaction be odd in the momentum, the analogue of the spin-orbit interaction for polaritons, the so-called TE-TM splitting, is even in the momentum.   We calculate and compare spin transport of   polariton, electron, and hole systems, in the diffusive regime of many scatterings.   After dimensional rescaling diffusive systems with spatially uniform particle densities have identical dynamics, regardless of the particle type.  Differences between the three particles appear in spatially non-uniform systems, with pumps at a specific localized point.  We consider both oscillating pumps and transient (delta-function) pumps.     In such systems each particle type produces  distinctive spin patterns.  The particles can be distinguished by their differing spatial multipole character, their response and resonances in a  perpendicular magnetic field, and their relative magnitude which is largest for electrons and weakest for holes.  These patterns are manifested both in response to unpolarized pumps which produce in-plane and perpendicular spin signals, and to polarized pumps where the spin precesses from in-plane to out-of-plane and vice versa.  These results will be useful for designing systems with large spin polarization signals, for identifying the dominant spin-orbit interaction and measuring subdominant terms in experimental devices, and for measuring the scattering time and the spin-orbit coupling's magnitude.
 \end{abstract}


\maketitle
 
 \section{Introduction}
Quantum particles are characterized not only by  their position, but also by additional quantum  numbers.  Among these quantities the electron's spin, a property   which has one of two values: $+ 1/2$, and $-1/2$, has tremendous promise for improvements in computing technology.  
If there is no magnetic field which breaks time reversal symmetry, then a 180 degree rotation reversing an electron's momentum must also reverse  the evolution of its spin.  Here we will explore diffusive spin transport of an alternate particle, which like electrons has a quantum number with two possible values, but which unlike electrons keeps the same dynamics when momentum is reversed.

The particle of interest is the cavity polariton, the  elementary excitation of semiconductor microcavities in the limit of strong coupling between light and the quantum well's excitons. Like electron spin, the polariton has a quantum number with two states: its polarization. 
Unlike electrons,  reversing  a polariton's momentum has no effect at all on its polarization.  To reverse the polarization's dynamics one must rotate the polariton by 90 degrees not 180 degrees, and thus interchange the $x$ and $y$ axis.    (For a review of polarization properties of polaritons see e.g. Ref.  \onlinecite{shelykh2010polariton}.) The consequences of this profound difference for polarization transmission under repeated scattering are at the heart of the current article.

The electron's single particle density matrix $\rho$ is characterized by charge density $N$ and spin densities $S_x, \, S_y, \,S_z$ and  can be represented as a vector
 \begin{eqnarray} 
 {\rho} &=& \begin{bmatrix} N, & S_x, & S_y, & S_z \end{bmatrix}.
 \end{eqnarray}
Like the electron,  the polariton also has a density matrix composed of its particle density $N,$  circular polarization density $S_z$, and linear polarization densities $S_x, \, S_y$ corresponding to $xy$ and diagonal in-plane polarization. In the ballistic regime, i.e without scattering, the polaritonic analog of spin-orbit coupling causes  the polarization's direction to precess, converting  linear $S_x, S_y$ polarization   to circular $S_z$ polarization  and vice versa.  For polaritons the axis of precession varies continuously over 360 degrees when the polariton momentum rotates by 180 degrees, unlike electrons where the axis of precession is locked to the momentum. 
Because of this angular structure, an  initial population of linearly polarized $S_x$ or $S_y$ polaritons will evolve into  into a four-lobed, or quadrupole, pattern of circular $S_z$ polarization.  In this celebrated effect, two lobes have positive circular polarization, two have negative polarization, and separating the four lobes is the Langbein cross where the initial linear polarization is unchanged.  \cite{PhysRevLett.95.136601, Leyder07, PhysRevB.75.075323, PhysRevB.79.125314, PhysRevB.80.165325, PhysRevLett.109.036404, PhysRevB.88.035311}  When a perpendicular magnetic field is added, the four-quadrant polarization pattern twists into spiral shapes. \cite{PhysRevB.88.035311, PhysRevLett.110.246403}

We will compare the polariton's dynamics under scattering to those of electrons,  and also to holes realized in zinc blende semiconductors, where the Fermi level is doped into the $J=3/2$ valence band rather than the conduction band.  In each of these cases we will restrict our analysis to two-dimensional systems analogous to the 2-D electron gas.    Like polaritons, electrons and holes can be written mathematically as two-state systems, and their dynamics are governed by  two-state Hamiltonians.  However, unlike polaritons,  they  possess  real spin and therefore must reverse their spin dynamics when their momentum is rotated by 180 degrees.  Moreover,  in holes (without strain or anisotropy \cite{PhysRevB.90.115306,PhysRevB.89.161307})  rotating the momentum by 60 degrees also reverses the spin dynamics.   This is a consequence of the crystal symmetry, in combination with the four-fold degeneracy of the valence band. \cite{PhysRevB.62.4245} Therefore an initial population of $S_x$ holes will evolve into a six-lobed, or sextupole, pattern of $S_z$ spin density.   Holes produce six lobes, polaritons produce four lobes, and electrons produce a simple dipole pattern. These qualitative differences in angular dependence correspond to precise differences in the two-state Hamiltonians governing electrons, polaritons, and holes.

In addition to the distinctive  spin-orbit splitting terms of each type of  particle, we will consider also the effect of a Zeeman splitting term caused by a external magnetic field oriented perpendicularly to the sample.   For polaritons this term may also be caused by variations in the polariton cavity such as in-plane strains inside the quantum well, or asymmetries in the direction of crystal growth.    \cite{shelykh2010polariton}  This external and tunable parameter is a useful probe of (pseudo)spin dynamics, and in polariton systems is known to twist the polarization pattern into a spiral.

Previous  polariton experiments have generally utilized $GaAs$ samples where the clean sample's low disorder density is further diminished by the polariton's large spatial extent, allowing polaritons to move ballistically, i.e. without scattering, on millimeter length scales. 
\cite{PhysRevLett.77.4792,PhysRevX.3.041015,PhysRevB.88.235314, steger2014slow,PhysRevB.88.245307}  
Ballistic  polariton transport has been studied both experimentally and theoretically in many papers.   \cite{PhysRevLett.95.136601, Leyder07, PhysRevB.75.075323, PhysRevB.79.125314, PhysRevB.80.165325, PhysRevLett.109.036404, PhysRevB.88.035311,PhysRevLett.110.246403} It is completely controlled by  (pseudo)spin precession:  As the polaritons move radially out from the polariton pump, their quadrupole pattern changes sign twice  for each precession length $l_{prcsn}$.   The result is that concentric circles of alternating sign are superimposed on the quadrupole pattern. \cite{PhysRevLett.109.036404}  Perfect spin polarization is maintained until the polaritons decay.

In this article we focus on a different scenario, the diffusive regime, where the scattering length is so short that  transport is governed by diffusion.      We assume that the experimental length and time scales  are long compared to the scattering scale. This is the typical case for experiments on 2-D electron and hole gases.  For polaritons, the diffusive regime can be obtained  by fabricating nanopillars  embedded in the polariton sample, similarly to optical systems where disorder has been induced by growing nanocolumns. \cite{sakai2014near, PhysRevLett.112.116402}    In the  diffusive case the particle's average displacement $r_0$ grows according to the diffusive law 
\begin{eqnarray}
r_0 &=& \sqrt{D \,t}
\end{eqnarray}
 where $D$ is the diffusion constant, unlike the ballistic regime where the rms value of $r_0$ grows linearly with time.  In diffusive samples the momentum becomes  randomized through elastic scattering, subject to the constraint that the energy remains equal to the pump energy $E_p$.  Therefore one calculates   the average dynamics and transport - averaging over the random disorder, and averaging over the randomized momentum.   Performing this averaging, and assuming a low-density, non-interacting system, we calculate the spin diffusion equations which govern the evolution of spin and particle densities.  Previous works on polaritons in the diffusive regime used a classical kinetic equation to calculate the equilibrium of steady-state spatially uniform systems. \cite{PhysRevB.77.165341,PhysRevB.82.085315} Here we calculate the dynamics of non-uniform systems using  standard perturbative techniques based in quantum mechanics which have been applied extensively to electron diffusion.
  
We will show that in the diffusive regime spatially uniform distributions of electrons, polaritons, and holes exhibit identical spin dynamics.  The essential reason is that the three particles differ only in their spin-orbit interaction's dependence on momentum, and momentum is randomized by scattering in the diffusive regime.   Differences between the spin diffusion equations of the three particles can be resolved only by breaking translational invariance, so we will focus entirely on non-uniform systems.

Previous articles deriving spin diffusion equations have often applied the equations to quasi-1-D geometries in steady-state, i.e. without temporal dynamics.  \cite{PhysRevLett.93.226602, PhysRevLett.105.066802}   In this case the diffusion equations simplify and can be solved analytically.  In this present article we focus instead on the richer and physically relevant case of  spatially localized pumps introducing particles into the device at one particular point.   Such pumps produce a rich set of interesting spin behaviors.  We will show that in the presence of a magnetic field, unpolarized pumps produce through scattering both in-plane and out-of-plane spin densities.  Polarized pumps do the same, and in addition exhibit spin precession between in-plane and out-of-plane spins.  These signals display spatial multipole patterns with the same angular structure as the spin-orbit coupling - a quadrupole structure in the case of polaritons. If there is a perpendicular magnetic field, the multipoles twist into spirals  as time progresses.  However, unlike ballistic transport, the spin pattern does not manifest concentric circles around the pump, since transport occurs according to a random walk rather than radial motion.

The structure of this paper is as follows.   In section \ref{sHamiltonian} we explain in detail the electron, polariton, and hole systems which can be realized experimentally, and realistic Hamiltonians for modeling these systems.  We also describe how to quantitatively compare diffusive transport of these disparate particles, by rescaling the dimensions and comparing dimensionless ratios.  In section \ref{sDensity} we mathematically prescribe the observables which are calculated in our paper - the density matrix which encodes spin and particle densities, and the response function for a spatially localized pump.  Next section \ref{sDiffusion} presents our analytical results: the diffusion equations which control the evolution of spin densities.   These diffusion equations are obtained using a standard technique which has been widely applied to  electron diffusion problems.  \cite{PhysRevB.70.155308,PhysRevB.81.125309,PhysRevB.71.121308, PhysRevB.72.075366, PhysRevB.74.195330,  PhysRevB.78.155302, PhysRevB.84.035318, PhysRevLett.105.066802}   Examination of the diffusion equations reveals the multipole pattern which will allow experimentalists to quickly determine the dominant spin-orbit coupling and carefully measure sub-dominant couplings.   Changing the spin-orbit splitting or the scattering time will cause the spatial multipole pattern to rotate.  We also show that in a perpendicular magnetic field polaritons and holes exhibit  a resonance that can be used to measure the spin-orbit strength.  The resonance is characterized by extinction of the  polarization produced in response to an unpolarized pump.   It occurs at numerically different magnetic field strengths for the two particles, and is sensitive to the spin splitting.    

The behavior of a real experimental device can not be determined unless the diffusion equations of section \ref{sDiffusion} are combined with initial conditions and solved.  Unlike previous works which analytically calculated steady-state quasi-1-D devices, in section \ref{sNumerical} we solve 2-D spin diffusion from a localized pump, using numerical techniques.   The problem/solution space is very rich: oscillating pumps vs. transient pumps, four types of spin input and response ($S_x, S_y, S_z$, and density $N$), various angular and radial profiles of the response, time dependence at both short and long times,  and an intricate dependence on the relative strengths of the  kinetic energy, spin-orbit splitting, Zeeman field, and scattering time. Surveying this complex solution space was one of the chief challenges of the present work. Section \ref{sNumerical} presents the most interesting features.  In many cases our results quantitatively confirm and illustrate the qualitative trends discussed in the previous section \ref{sDiffusion}.  However Section  \ref{sNumerical} also gives qualitatively new information: we show that the time evolution breaks naturally into two regimes, before and after loss of memory of the pump polarization. Spin observables can reach quite large values in the first regime, and can persist well beyond the memory loss.  We also  examine the evolution into spiral structures which will allow measurement of the scattering time, and  determine the parameters which maximize the spin polarization.    Lastly Section \ref{sDiscussion} synthesizes the major differences between electrons, polaritons, and holes.  Several appendices present mathematical details of our calculations.

\section{The Microscopic Hamiltonians for Electrons, Polaritons, and Holes\label{sHamiltonian}}
We now introduce a  realistic Hamiltonian which  models polaritons in a GaAs quantum well, as well as electronic 2DEGs and hole-doped systems.   The spin-orbit coupling varies substantially in these three systems.  It is the linear Rashba coupling for electrons, a quadratic  coupling for polaritons, and a cubic  coupling for holes. We write the Hamiltonian for all three cases in a unified notation with $N=1$ for electrons, $N=2$ for polaritons, and $N=3$ for holes:
\begin{eqnarray}
\label{NHamiltonian}
H^N &=& \frac{\hbar^2 k^2}{2m} + b_z \sigma_z+ \Delta\begin{bmatrix}0& (k_x - \imath k_y)^N \\ (k_x + \imath k_y)^N & 0\end{bmatrix} 
\nonumber \\ 
 &=&\frac{\hbar^2 k^2}{2m} + C\; E_{prcsn} \sigma_z
\nonumber \\
&+& S\; E_{prcsn} k_F^{-N} \begin{bmatrix}0& (k_x - \imath k_y)^N \\ (k_x + \imath k_y)^N & 0\end{bmatrix},
\\ \nonumber 
  S &=& \sin \theta_B = \frac{\Delta k_F^N}{E_{prcsn}}, \; C = \cos \theta_B =  \frac{b_z}{E_{prcsn}}, 
  \\ \nonumber 
   E_{prcsn} & = & \sqrt{(\Delta k_F^N)^2 + b_z^2}
\end{eqnarray} 
where $\vec{\sigma} = \left[ \sigma_x, \sigma_y, \sigma_z \right]$ are the Pauli sigma matrices, $\vec{k}$ is the wave vector, and $k_F$ is the characteristic wave-vector determined by the energy $E$.   For convenience we have expressed the relative strength of the Zeeman term and the spin-orbit coupling with the angle $\theta_B = \arctan(\Delta k_F^N / b_z)$, a sine $S = \sin \theta_B $,  and a cosine $C = \cos \theta_B$.      $2 E_{prcsn} =2 \sqrt{(\Delta k_F^N)^2 + b^2}$ is the  splitting between energy levels.

 This is a non-interacting Hamiltonian, which works well  for polaritons in the limit of low polariton density, where polariton interactions are weak and can be neglected.  In this limit the polariton energy $E_p$ - the in-plane kinetic energy in the lower polariton branch - is the same as the energy of the pump photons used to excite the polaritons.  Moreover the polaritons' characteristic wavelength $k_F$ is  the magnitude of $\vec{k}$ on the contour in momentum space  defined by $E(\vec{k}) = E_p$. (This contour is  the elastic scattering circle for polaritons, or the Fermi surface for electrons and holes.)  Note that in the regime of strong pumps nonlinear effects may qualitatively change the texture of the spin patterns \cite{PhysRevLett.110.016404,PhysRevLett.110.035303}.  The non-interacting picture is also appropriate for electrons in 2DEGs and for hole-doped semiconductors, with the caveat that here $E_p$ is the Fermi energy $E_p$, and is controlled by doping, gating, etc.  
 
  Our  $N=2$ polariton Hamiltonian is accurate  when the polariton energy $E_p$ is a fraction of  Rabi splitting $\Omega_R$, which is the energy scale  of the exciton - photon coupling that produces the polaritons.  In $GaAs$ the Rabi splitting is near the value  $\Omega_R = 6 meV$ seen in Refs. \onlinecite{PhysRevB.53.R10469, PhysRevLett.73.2043, PhysRevLett.89.077402, PhysRevX.3.041015,PhysRevB.88.235314, steger2014slow}, so our Hamiltonian is valid in the range $E_p = \left[0, 2 \right] \, meV$. 
 In this article we will fix $E_p$ at $2.0 \,meV$; decreasing this value only rescales the scattering length and time, as well as the polariton wavelength $l_p$ determined by the pump energy.     We also choose the  mass $m = 5 \times 10^{-5} m_e$, as reported in References \onlinecite{PhysRevX.3.041015,PhysRevB.88.235314, steger2014slow}. This value is typical for polaritons in $GaAs$ quantum wells, and is determined by the band gap in $GaAs$.  

The polarization evolution (or spin evolution for electrons and holes) is determined by the second and third terms in $H$.  The  second term  is a Zeeman splitting depending on the polarization (or spin) quantum number.  Here we assume that this term is oriented along the $\hat{z}$ axis, perpendicularly to the sample, so it produces only in-plane precession.  

The third term gives the momentum-dependent spin-orbit coupling, or the TE-TM splitting in the case of the cavity polariton.  In polariton systems symmetry arguments require that this term be even in the momentum, which is a very distinctive signature of polaritons.   In contrast, in electron and hole systems  time reversal symmetry requires that this term be odd in the momentum, as long as there is no magnetic field or magnetic impurities.  The $N=1$ linear term we use here for electrons is just the Rashba spin-orbit interaction. 

The principal source of the quadratic polariton TE-TM splitting is the confined photons' sensitivity to the mirrors of the cavity and to the angle of the incident light. \cite{shelykh2010polariton}   Studies of  the TE-TM term  and of the Zeeman term (with a modest $5$ Tesla field) have shown that both terms can reach or exceed $0.2 \, meV,$  so that they may be comparable to the total polariton energy, and spin-orbit effects can be very large.  \cite{PhysRevB.53.R10469,PhysRevB.59.5082,PhysRevB.67.195321,PhysRevLett.92.017401,PhysRevB.84.165325,schneider2013electrically}

\subsection{How to Compare Electrons, Polaritons, and Holes}
For the purpose of comparison between  uncharged polaritons and charged electrons and holes, we  neglect the electrical potential $\Phi$ and magnetic gauge potential $\vec{A}$ acting on charges. The  gauge potential  $\vec{A}$ can be neglected if the cyclotron radius is larger than both the scattering length and the dephasing length, which is easily achieved if the sample is not very clean and the temperature is high enough to extinguish weak localization effects.   Turning to the electrical potential, a gradient in this potential will cause a net drift of the charges, which  will act in addition to the diffusive dynamics considered here.  This drift can be be minimized by decreasing the potential gradient.  

In the preceding discussion of the Hamiltonian we have taken care to use realistic values for the polariton mass $m$, pump energy $E_p$, and spin-orbit coupling strength.   However the final results of our calculations are determined only by length and time scales and dimensionless ratios formed from combinations of these parameters.  We can compare polariton results directly to electron and hole results by rescaling the dimensions to match each other.   Here are the physically relevant scales:

\textbf{The  Wavelength $l_p$ and Velocity $v_p$.} The mass $m$ and the pump energy $E_p$ (or Fermi energy for electrons and holes) determine  the  wavelength 
\begin{eqnarray}
l_p &=& h /\sqrt{2 m E_p},
\end{eqnarray}
 which is the fundamental length scale governing transport and dynamics. They also determine the polariton's velocity in ballistic flight, 
 \begin{eqnarray}
 v_p &=&  \sqrt{2 E_p/m},
 \end{eqnarray}
  which conveniently allows conversion of times and energies to lengths, and vice versa.  At the polariton energy $E_p = 2.0 \, meV$ 
the polariton wavelength and velocity are respectively
$l_p =4 \, \mu m$ and $v_p = 4  \, \mu m / ps$.  

\textbf{The Precession Length $l_{prcsn}$.}  Taken together, the Zeeman and spin-orbit terms determine the  spin (or polarization) precession length 
\begin{eqnarray} 
l_{prcsn} &=& \hbar v_p / E_{prcsn},
\end{eqnarray}  where $E_{prcsn} = \sqrt{(\Delta k_F^N)^2 + b^2}$.    We assume that the spin-orbit term is small compared to the kinetic energy $p^2 /2m$, or equivalently, that 
the precession length $l_{prcsn}$ is large compared to  $l_p =4 \, \mu m$.

\textbf{The Scattering Length $l$} scales with
\begin{eqnarray} 
l \propto \frac{(\hbar v_F)^2}{\langle V^2 \rangle\, \upsilon^2 \, k_F}\
\end{eqnarray}
where $\langle V^2 \rangle$ the second moment of the disorder potential and $\upsilon$ is the characteristic length scale of the impurities.  In the diffusive regime only $l$ itself determines the system's evolution, not $n$ or $\langle V^2 \rangle$.    We do require that the disorder be weak according to the Ioffe-Regel criterion; i.e. that $\chi = l_p / l \pi \ll 1$.  Otherwise the diffusion picture breaks down and Anderson-localized states may be observed.  To ensure diffusive transport in our polariton system with $l_p = 4 \, \mu m$, we set $l = 16 \, \mu m$.  The scattering time is $\tau = l / v_p =  4  \, ps$.

\textbf{The Polariton Decay Length $l_{decay}$.} Unlike electron and hole systems where  charge is conserved, polariton number generically decreases with time as photons escape through the Bragg mirrors of the quantum cavity.  This decay is determined by the polariton lifetime  $\tau_{decay}$, which also determines the maximum distance  $l_{decay}$ that polaritons travel before decaying. Under diffusive motion this distance is 
\begin{eqnarray}
l_{decay} &=& \sqrt{D \tau_{decay}}
\end{eqnarray}
 where $D = l v_p / 2$ is the 2-D diffusion constant.
  Diffusion can be observed only if the decay length $l_{decay}$ substantially exceeds the scattering  length $l = 16\, \mu m$.  Therefore we assume a relatively large value  $l_{decay} = 80\, \mu m$, corresponding to $\tau_{decay} = 200 \, ps = 50 \tau$.  The (pseudo)spin dynamics of greatest interest occur at shorter time scales of order $10\tau$, so our numerical results on spin are largely independent of the decay time.  For electrons and holes the decay time is infinite.

  These numerical values of the polariton decay length and time are experimentally accessible.  Polariton cavities far larger than $ l_{decay} = 80\, \mu m$ are  routinely fabricated in  present experiments.  Improved mirrors have recently  increased the lifetime from a few $ps$ to $180$ ps,  so $200 ps$ is also quite feasible.  \cite{PhysRevX.3.041015,PhysRevB.88.235314, steger2014slow}

Given these length scales, electrons, holes, and polaritons can be compared directly to each other after   positions are normalized by the scattering length $l$ and times are normalized by $\tau$.  These are the natural length and time scales of a diffusive system.  After performing this normalization, only four dimensionless quantities determine the (pseudo)spin dynamics:
\begin{itemize}
\item The dimensionless disorder strength  $\chi = l_p / l \pi= \hbar / E_p \tau  \ll 1 $.
\item $\theta_B$, which controls the relative strength of the Zeeman and spin-orbit terms.
\item The dimensionless diffusion length, $r_0/l= \sqrt{ t/2 \tau}$, which tracks the diffusive spreading of the initial particles.
\item The dimensionless energy splitting between the spin up and down states $\zeta  = l/l_{prcsn} = E_{prcsn} \tau / \hbar$ which  determines the dominant source of spin relaxation. \cite{wenk2010spin, wu2010spin, PhysRev.96.266, Dyakonov72, Yafet63} 
\end{itemize}
  This last parameter is key to the spin dynamics.  $\zeta \gg 1/2$ puts the system in the Elliott-Yafet (EY) regime, where the (pseudo)spin precesses many times before scattering and momentum is tightly coupled to spin.  In the EY regime the spin is randomized at every scattering event, so its relaxation time is  close to the scattering time $\tau$.    In contrast, $\zeta \ll 1/2$ puts the system  in the D'yakonov-Perel' (DP) regime  where the (pseudo)spin precession length $l_{prcsn}$ is long compared to $l$, and the energy splitting  is small compared to the scattering energy. In this case  precession lasts much longer than the scattering time, and the spin relaxation length  scales with $l_{prcsn}$.   

As always, we assume that  $l_p / l_{prcsn}  =  \pi \zeta \chi$ is small, so that the spin splitting is small compared to the kinetic energy $p^2/2m$.

\section{\label{sDensity}The Density Matrix and the Response Function}

We now begin developing the equations describing (pseudo)spin dynamics in the diffusive regime, where the particle scatters randomly many times, and where we must average over disorder in order to calculate transport properties.  Disorder-averaged diffusive transport is qualitatively different from transport in the ballistic regime, where all transport information is contained in the wave-function $|\psi\rangle$.  In  the diffusive regime scattering randomizes $|\psi\rangle$'s phase and therefore its disorder average  $\langle |\psi \rangle \rangle_\epsilon$ averages to zero, losing all transport information.   Study of the disorder-averaged wave-function alone is unable to describe transport, or even probability conservation, in the diffusive regime.  

A correct description of diffusive transport begins with the single-particle density matrix which for the case of a pure state reads:
\begin{eqnarray}
\rho(t) &= & |\psi (t) \rangle \otimes \langle \psi(t)|
\\ \nonumber
\rho(t, \vec{x}, \acute{\vec{x}}) &=& \langle \vec{x} | \psi(t) \rangle \; \langle \psi(t) | \acute{\vec{x}}\rangle
\end{eqnarray}
The density matrix encodes complete information about the probability density $I(\vec{x})$ in its local $\vec{x} = \acute{\vec{x}}$ values:
\begin{eqnarray}
I(t,\vec{x}) &=& {Tr}(\rho(t, \vec{x}, \vec{x}))
\end{eqnarray}
Here the trace is performed over $\rho$'s spin indices. $I(\vec{x})$ is proportional to the charge density in electron and hole systems, and to the particle density in polariton systems. Similarly, the (pseudo)spin densities are also encoded in $\rho$'s local values:
\begin{eqnarray}
S_x(t,\vec{x}) &=& {Tr}(\rho(t,\vec{x}, \vec{x}) \, \sigma_x)
\nonumber \\
S_y(t,\vec{x}) &=& {Tr}(\rho(t,\vec{x}, \vec{x}) \, \sigma_y)
\nonumber \\
S_z(t,\vec{x}) &=& {Tr}(\rho(t,\vec{x}, \vec{x}) \, \sigma_z)
\end{eqnarray}
These local quantities survive the disorder average because of the cancellation of phases between $|\psi \rangle$ and $\langle \psi |$.  Nonlocal $\vec{x} \neq \acute{\vec{x}}$ values of the density matrix decay exponentially when $\vec{x} - \acute{\vec{x}}$ is longer than the scattering length if we are outside the polariton condensation regime \cite{PhysRevLett.113.203902}.  We confine our attention to the local quantities  $I(t,\vec{x}), \, S_x(t,\vec{x}), \, S_y(t,\vec{x}), \, S_z(t,\vec{x})$, which  are the proper starting point for studying diffusive transport.  We write these as a vector:
\begin{eqnarray}
(\rho(t,\vec{x}))^T & = & \{ I(t,\vec{x}), \, S_x(t,\vec{x}), \, S_y(t,\vec{x}), \, S_z(t,\vec{x}) \}\;\;\;
\end{eqnarray}
It is also convenient to study the spatial Fourier transform of the density matrix, because  random disorder after averaging does not break the (average) translational invariance:
\begin{eqnarray}
(\rho(t,\vec{q}))^T & = & \{ I(t,\vec{q}), \, S_x(t,\vec{q}), \, S_y(t,\vec{q}), \, S_z(t,\vec{q}) \}
\end{eqnarray}
When studying transport with an oscillating pump we will also study the temporal Fourier transform
\begin{eqnarray}
(\rho(\omega,\vec{x}))^T & = & \int {d\omega} \, \exp(\imath \omega t) \, (\rho(t,\vec{x}))^T
\end{eqnarray}

It is often useful to compare the spin densities $S_i$ to the probability density $I$. We define the (pseudo)spin degrees
\begin{eqnarray}
s_x &=& S_x/I, \; s_y = S_y/I, \; s_z = S_z/I
\end{eqnarray}
These degrees can never be larger than $1$ or smaller than $-1$.   At one extreme, the density matrix may be entirely unpolarized, in which case it is proportional to $\vec{\rho} = \{1, 0, 0, 0\}$ and the spin degrees are all zero. At the other extreme it may be completely  polarized, with $s_x^2 + s_y^2 + s_z^2 = 1$.

\subsection{ Response Functions}
Our goal is to obtain and study a response function $\Phi(t)$ which evolves the density matrix forward in time:
\begin{eqnarray}
\label{ConvolutionEqn}
\rho(t, \vec{x}) &=& \int\, {d\acute{\vec{x}}}\; \Phi(t, \vec{x} - \acute{\vec{x}})\, \rho(t=0, \acute{\vec{x}})
\end{eqnarray}
$\Phi(t, \vec{x}, \acute{x})$ is a $4\times 4$ matrix, because it computes the charge and spin densities at time $t$ as function of the charge and spin densities at $t=0$.  The response function is also a linear operator, which is a direct consequence of our working in the noninteracting regime.    Therefore we will study only the response $\Phi(t, \vec{x})$ to a $t=0$ delta function input $\vec{\rho}_0  \, \delta^2(\vec{x})$ at the origin.  The response to any input may be reconstructed from $\Phi(t, \vec{x})$ by convolving the response function over the $t=0$ density matrix according to equation \ref{ConvolutionEqn}.

We will also calculate the temporal Fourier transform  $\Phi(\omega, \vec{x})$ which describes the density matrix produced in response to a pump $\rho_0$ oscillating with period $T = 2 \pi / \omega$:
\begin{eqnarray}
\label{ConvolutionEqnOmega}
\Phi(\omega, \vec{x}) &=& \int {dt}\; \exp(\imath \omega t) \; \Phi(t, \vec{x})
\nonumber \\
\rho(\omega, \vec{x}) &=& \int\, {d\acute{\vec{x}}}\; \Phi(\omega, \vec{x} - \acute{\vec{x}})\, \rho_0(\omega, \acute{\vec{x}})
\end{eqnarray}
This frequency response function contains both information about the response amplitude, and also information about the response's phase relative to the pump.  Therefore it is in general complex, with both real and imaginary components.  In this paper we will study only the magnitude of the response, not its phase information.

Because the system is translationally invariant after averaging over disorder, it is convenient to Fourier transform both the time response function $\Phi(t, \vec{x})$ and frequency response function $\Phi(\omega, \vec{x})$ to momentum space: 
\begin{eqnarray}
\label{SpatialFT}
\rho(t, \vec{x}) &=& \int\, \frac{d{\vec{q}}}{(2 \pi)^D}\; \exp(\imath \vec{q} \cdot \vec{x}) \; \Phi(t, \vec{q})\, \rho(t=0, {\vec{q}})\;\;\;
\end{eqnarray}


  We  determine the response functions  using standard methods from the diagrammatic technique for disordered systems \cite{Hikami80}.    We neglect effects from quantum interference and adopt a purely classical picture of particle diffusion through the disordered system. 
      We model scattering with a  "white noise" disorder potential  which does not alter the (pseudo)spin 
   \begin{eqnarray}
   V &=& \begin{bmatrix} 1 &  0 \\ 0 & 1 \end{bmatrix} u(\vec{r}), \, \langle u(\vec{r}) u(\acute{\vec{r}}) \rangle =  (2 \pi \nu \tau / \hbar)^{-1} \delta(\vec{r} - \acute{\vec{r}})\;\;\;\;\;
   \end{eqnarray}
    where   the density of states $\nu(E_p)$ at pump energy $E_p$ is 
    \begin{eqnarray}
    \nu(E_p) &=& \sum_s \int \frac{d\vec{k}}{(2\pi)^D}\, \delta(E_p -  E(s, \vec{k}))
    \end{eqnarray}  
    and $E(s, \vec{k})$ are the two eigenvalues of the Hamiltonian at wave-vector $\vec{k}$.   We assume that this equilibrium value of $\rho$ is unpolarized, which just means that $E_p > 2 E_{prcsn}$.  With this  white noise disorder every scattering event causes the particle to lose its memory of its previous momentum, which becomes evenly distributed on the elastic scattering circle.   Under these assumptions the time response function can be written as an exponential:
    \begin{eqnarray} \label{PhiT}
    \Phi(t, \vec{q}) &=& \exp(\frac{t}{\tau} \, D^{-1}(\vec{q})), \; t > 0
    \end{eqnarray}
    $D^{-1}(\vec{q}) = 1 - I_{ij}$ is the inverse of an operator called the diffuson.  It contains information about a single scattering event which is represented by the scattering operator $I_{ij}$.  The response function is obtained by exponentiating $D^{-1}$,  describing the results of repeated scatterings. As a consequence of this exponential form, the frequency response function is
    \begin{eqnarray} \label{PhiOmega}
    \Phi(\omega, \vec{q}) &=& \frac{1} {\imath \omega + \tau^{-1} D^{-1}(\vec{q})}
    \end{eqnarray}
    Equivalently to equations \ref{PhiT} and \ref{PhiOmega}, we can also write the spin diffusion equation
 \begin{eqnarray} 
    \frac{\partial}{\partial t}\rho(t, \vec{q}) &=&  D^{-1}(\vec{q})  \,\rho(t, \vec{q}).
    \end{eqnarray}
This reveals that $D^{-1}$ is the diffusion operator controlling evolution of the density matrix.  Before solving the diffusion equation one must first supplement it with the starting state $\rho(t=0)$.

 \begin{figure}[]
\begin{center}
  \includegraphics[width=8cm, height=3cm, bb=0 0 480 220]{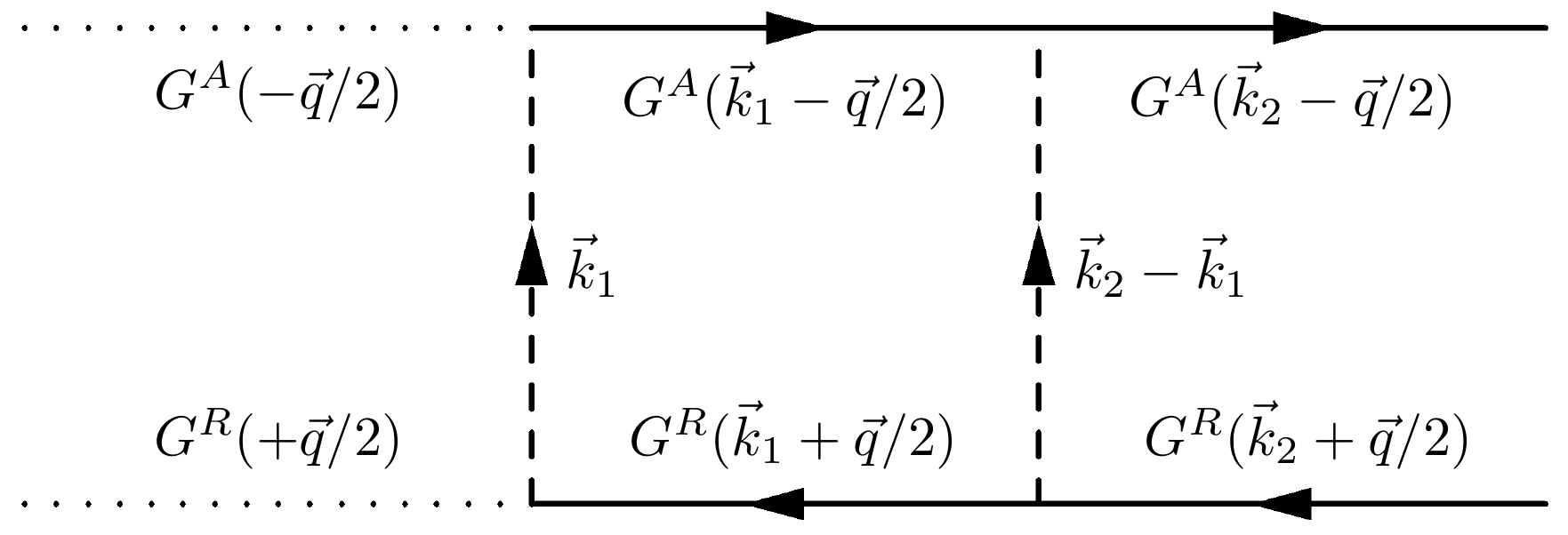} 
    \end{center}
    \caption{
Diagram illustrating the scattering that produces  diffusion.  Here two scattering events are shown. $G^A$ and $G^R$  describe time evolution of the polariton $\psi$ and its complex conjugate  $\psi^\dagger$.  Each scattering event causes correlations between $G^A$ and $G^R$ and is shown as a dashed line connecting the two. } 
    \label{fig:JointScattering}
\end{figure}

In the next section we will derive the diffusion operator $D^{-1}(\vec{q})$ analytically.  Calculation of the response functions requires numerical calculations - either exponentiation  (Equation \ref{PhiT}) or inversion (Equation \ref{PhiOmega}) - and we will perform this numerical work in Section \ref{sNumerical}.  

\section{\label{sDiffusion}Analytical Calculation of Diffusive Transport}

 In order to calculate the diffusion operator $D^{-1}(\vec{q})$, one must first calculate the scattering operator $I_{ij}$.   In each scattering event  the  wave-function $|\psi\rangle$ and its conjugate $\langle \psi |$ move together, scattering in unison.   Two scattering events are  pictured in Figure ~\ref{fig:JointScattering}.   The scattering operator $I_{ij}$ is given by the integral
   \begin{eqnarray}
   I_{ij} & = & \frac{\hbar}{4 \pi \nu \tau} \int d\vec{k} \, {Tr}(\, G^A( \vec{k} - \vec{q}/2, E_p)   \sigma_i  
\nonumber \\
& \times & \, G^R(  \vec{k} + \vec{q}/2,  E_p ) \, \sigma_j ) 
\label{JointScatteringOperator}
\end{eqnarray}
Here $G^A$ and $G^R$  are the disorder-averaged single-particle Green's functions, and their spectral representation is given in equation \ref{GreensFunctionSpectralRep} in Appendix \ref{AppDiscrepancy}.   The trace is taken over the spin indices of $G^A, G^R, \sigma_i,$ and $\sigma_j$, which are all $2 \times 2$ matrices in (pseudo)spin space.    $\vec{q}$ is the diffuson wave-vector.  Further details of the integration of equation \ref{JointScatteringOperator} are given in equations  \ref{DiffusonDetails1} and \ref{DiffusonDetails2} of   Appendix \ref{AppDiscrepancy}.

Equation \ref{JointScatteringOperator} for the scattering operator, and $D^{-1}(\vec{q}) = 1 - I_{ij}$  for the diffuson,  are a well established and widely adopted formalism for determining spin and charge diffusion.  \cite{PhysRevB.70.155308,PhysRevB.81.125309,PhysRevB.71.121308, PhysRevB.72.075366, PhysRevB.74.195330,  PhysRevB.78.155302, PhysRevB.84.035318, PhysRevLett.105.066802}     This formalism has  a strong physical motivation expressed in the perturbative diagram in Figure \ref{fig:JointScattering},  can be derived from the Keldysh equations for nonequilibrium conduction, and is equivalent to Kubo linear response theory.  \cite{PhysRevB.75.155335}  In the present work we use this formalism to study electrons, polaritons, and holes, in a magnetic field, to all orders of the spin splitting parameter $\zeta$.   We alsos have checked that our results coincide with those of Reference  \onlinecite{PhysRevLett.93.226602}, which is also based on the Keldysh equations, but uses a different strategy for solving them iteratively.   
 
As seen from the Fourier transform in equation \ref{SpatialFT}, there is a direct mapping from the momentum $\vec{q}$ to position space.  Precisely, every power of $\imath q_x$ corresponds to a $\partial/\partial_x$ spatial derivative, every power of $\imath q_y$ corresponds to a $\partial/\partial_y$, and $\imath \vec{q}$ corresponds to $\vec{\nabla}$.  Because we are interested only in length scales that are long compared to the scattering length $l$, we perform a Taylor series expansion of the scattering operator $I_{ij}$ in powers of $l\vec{q}  $ (or equivalently in powers of  $l \vec{\nabla}$) and terminate the expansion at second order.  This approximation is justified for calculating observables at scales larger than $l$.  It guarantees that the inverse diffuson contains only terms with at most two spatial derivatives, which is appropriate for describing particle diffusion under random scattering. \cite{PhysRevB.70.155308,PhysRevLett.93.226602} For the spin-charge coupling in holes we make an exception to this procedure and keep cubic $(lq)^3$ terms, since at quadratic order this coupling is zero, as is well known from previous studies.   \cite{PhysRevB.74.195330, PhysRevB.76.073314, PhysRevB.74.193316}

As mentioned before, we assume that disorder is weak, far from the localized phase, so that the dimensionless disorder strength $\chi = l_p / \pi l=\hbar / E_p \tau$ is small.  Therefore we expand $I_{ij}$ in a Taylor series  in powers of $\chi$ and truncate at first order.    We neglect all powers of $\chi $ that do not occur in combination with $l$, $\tau$, or $\zeta$. 

Our spin-orbit term has the special property that the spin splitting $2 E_{prcsn} = 2 \sqrt{(\Delta k_F^N)^2 + b_z^2}$ is rotationally invariant.  As a result we are able to analytically calculate the scattering operator $I_{ij}$ to all orders in the dimensionless energy splitting $\zeta = l/ l_{prcsn}$.


\subsection{The diffuson in the EY and DP regimes}

After Taylor expanding equation \ref{JointScatteringOperator} and integrating over the elastic scattering circle, we find that the inverse diffuson $D^{-1}$ decomposes into four parts: a momentum independent part $\lambda$ which gives the  spin (or polarization) relaxation rates and the  precession term, a diffusion term $\Delta$ which includes the  coupling between charge/number density and circular $S_z$ polarization, a momentum-dependent part $\kappa_{0l}$ which gives the coupling between  density and linear $S_x,S_y$ polarization, and a momentum-dependent part $\kappa_{lc}$ which gives the coupling  between linear $S_x,S_y$ polarization and circular $S_z$ polarization.  
Our decomposition reads:
\begin{equation} \label{DiffusonComponents}
 D^{-1} = -(\lambda - \Delta+ \kappa_{0l} + \kappa_{lc} + \tau/\tau_{decay}) 
\end{equation}
We present the four terms both in the  Elliott-Yafet limit where  $\zeta \gg 1/2$, and also in the D'yakonov-Perel' limit where  $\zeta \ll 1/2$.  In Appendix \ref{AppendixFullExpressions} we present the more complicated expressions which interpolate between these limits. Appendix \ref{AppDiscrepancy} shows gives the details of our derivation of $\kappa_{0l}$ and also of the coupling between particle density and $S_z$ polarization.

Spatially uniform spin distributions are governed by the mass and  precession matrix.  Very remarkably, this term is the same for electron, polariton, and hole-doped systems, regardless of $N = 1,2,3$.  It is:
\begin{eqnarray} \label{DiffusonA}
\lambda^{DP} &=&  \begin{bmatrix} 0 & 0 & 0 & 0 \\ 0 &2 \zeta^2(1 + C^2 )  &   2\zeta C & 0 \\ 0 & -  2\zeta C &   2\zeta^2 (1 + C^2) & 0 \\ 0 & 0 & 0 &4 \zeta^2 S^2   \end{bmatrix}, \;
\nonumber \\
 \lambda^{EY} &=& \begin{bmatrix} 0 & 0 & 0 &0\\ 0 &  (1 + C^2)/2 & 0& 0 \\ 0 & 0 &  (1 + C^2) / 2& 0 \\ 0 & 0 &  0 & S^2  \end{bmatrix} 
\end{eqnarray}
where the $C$ and $S$ parameters are defined in Eq. \ref{NHamiltonian}.  The on-diagonal terms  in $\lambda$ give the (pseudo)spin relaxation rates, which are the inverses of the spin lifetimes.  The upper left entry is the inverse lifetime of the charge/number density, and is zero because elastic scattering does not change the polariton number or the electron or hole charge. In the EY $\zeta \gg 1/2$ regime the lifetimes are proportional to the scattering time $\tau = 4 ps$, while in the DP $\zeta \ll 1/2$ regime they scale quadratically with $\zeta^{-1}$; $\tau_{xy},\tau_z \propto \tau / \zeta^2 \propto \tau^{-1}$.  

The off-diagonal terms of $\lambda$ (zero in the EY regime)  describe the usual precession of the  in-plane polarization around the magnetic field, which is oriented perpendicularly to the sample, along the $z$ axis.   They render the diffuson non-Hermitian, so that its eigenvalues are complex, as required for describing precession. In the DP regime the polarization precesses many - $O(\zeta^{-1})$ - times before it decays.  In contrast, in the EY regime polarization precession is not visible at diffusive scales, because each scattering randomizes the polarization. 

We turn to $\Delta$, the part of the diffuson which is proportional to $q^2$ and therefore describes diffusion.  Since it is zero at zero momentum $q=0$,  it has no effect on spatially uniform  distributions.  \begin{eqnarray} \label{DiffusonB}
-\Delta^{DP}&=&   \frac{(ql)^2}{2}  \begin{bmatrix}1 &0  & 0 & d_{0z} \\  0  & 1 & - 6 \zeta C  & 0 \\  0& 6 \zeta C  & 1 & 0 \\   d_{0z} & 0 & 0 & 1 \end{bmatrix}, \;
\nonumber \\
-\Delta^{EY} &=& \frac{(ql)^2}{2}\begin{bmatrix} 1 & 0 & 0 & d_{0z} \\ 0 &   S^2/2 & 0& 0 \\ 0 & 0 &   S^2/2 & 0 \\ d_{0z} & 0 & 0  & C^2 \end{bmatrix} 
\nonumber \\
  d_{0z,N} &=&  -\chi \zeta C +  \chi \zeta C S^2 \, N
  \nonumber \\
  &-& \chi\zeta     / (1+4 \zeta^2)^2\;     C S^2   \,N
\end{eqnarray}

 The most interesting aspect of this diffusion operator is $d_{0z}$, the coupling which controls the generation of  $S_z$ circular polarization from polariton density  (or electron or hole charge).    This coupling implies that from an initial population of unpolarized polaritons,  diffusive scattering in a magnetic field will produce an $S_z$ signal perpendicular to the plane, with a conversion efficiency controlled by $\chi \zeta = l_p / \pi l_{prcsn} $.  The resulting $S_z$ signal is rotationally symmetric in real space.   The  $d_{0z}$ coupling describes diffusive generation  of $S_z$ polarization inside the structure of the diffuson $D^{-1}$ itself.  It is not the same as the external Zeeman potential introduced in  Ref. \onlinecite{PhysRevB.70.155308}, which does not couple $S_z$ to density, and is not part of the diffuson. The  $d_{0z}$ coupling requires inclusion of the Zeeman splitting while computing $D^{-1}$; it is zero when $b_z = 0$, and therefore is omitted  in some previous works on diffusion with a Rasha term.   \cite{PhysRevLett.93.226602,PhysRevB.70.155308}

$d_{0z}$'s magnitude $\chi  \zeta (ql)^2$ is the same  for electrons, polaritons, and holes, but its dependence on $\theta_B$, the balance between Zeeman and spin-orbit terms, differs  for the three particles. We focus on the EY regime where the last term in $d_{0z}$ can be neglected.   In an electron system with linear spin-orbit coupling $d_{0z}$ does not depend on the spin-orbit strength $\Delta$ at all.  Interestingly, in polariton systems  if the magnetic field strength $|b_z| = |\Delta k_F^2|$ is tuned to exact resonance with the spin-orbit strength, then $d_{0z}$ goes to zero and changes sign. 
In hole-doped systems with cubic coupling the same resonance is still present, but is shifted to $|b_z| = \sqrt{2} |\Delta k_F^3|$.    This resonant behavior is interesting both because it is different for electrons, polaritons, and holes, and also because it may be useful for measuring the spin-orbit strength.

The diffusive component $\Delta$ contains  an additional term $\Delta^{1}$ which is proportional to $(ql)^{2N}$.  Interestingly, it breaks rotational symmetry, mirroring the spin-orbit coupling.  In  electron systems with a linear Rashba spin-orbit coupling, $\Delta^{1}$  is:
\begin{eqnarray} \label{DiffusonC}
\Delta^{1} &=& - e \frac{(ql)^2}{2}  \begin{bmatrix}0 &0  & 0 &0 \\  0  & \cos(2 \theta_q) & \sin(2 \theta_q)  & 0 \\  0& \sin(2 \theta_q)  & -\cos(2 \theta_q) & 0 \\   0 & 0 & 0 & 0 \end{bmatrix}, \;
\nonumber \\
 e &=& \frac{ 2  \zeta^2 S^2 (3 + 6 \zeta^2 + 8 \zeta^4) }{(1 + 4 \zeta^2)^3} 
\end{eqnarray}
In the DP $\zeta \ll 1/2$ regime this term is proportional to $\zeta^2$, which is a small contribution to $\Delta^{DP}$.  However in the EY $\zeta \gg 1/2$ regime where the spin-orbit coupling is strong $\Delta^1$ is proportional to $S^2/4$, and induces a quadrupole angular dependence in the diffusion. A similar anisotropic diffusion term has been found for spin conduction on the surface of a 3-D TI, which because of its strong spin-orbit coupling is in the EY regime. \cite{PhysRevLett.105.066802, PhysRevB.85.205303}  More generally, in polaritons $\Delta^1$ scales with $(ql)^4$, and in holes it scales with $(ql)^6$.    Because in polaritons and holes $\Delta^1$ is parametrically smaller than the  $(ql)^2$ part of $\Delta$, we neglect it in these systems, and retain it only for electrons.

Unlike Ref. \onlinecite{PhysRevLett.93.226602}'s treatment of the Rashba coupling which leaves the diffusion term constant at its $\zeta = 0$ value where the spin splitting is very small, our result for $\Delta$  keeps all orders of $\zeta$.  This allows spin or pseudospin  to diffuse at a rate different from that of charge or number density, a possibility discussed in references \onlinecite{pramanik2008inequality, PhysRevLett.76.3794, PhysRevB.79.115321}. 
While the  diffusion term  $\frac{(ql)^2}{2}$ controlling the number/charge density is insensitive to $\zeta$, the other terms in $\Delta$ are reduced by the spin-orbit coupling.  For instance, the $S_z$ diffusion term (expanded to second order in $\zeta$) is  reduced to $\frac{(ql)^2}{2}(1 - 24 \zeta^2)$.   
This implies that spatial inhomogeneities of number/charge density are smoothed more quickly than spatial inhomogeneities of (pseudo)spin.

We find also a coupling which generates   $S_x, \, S_y$ linear polarization from polariton number density  (or electron or hole charge density):
\begin{eqnarray}
\label{densitylinear}
\kappa_{0l}&=& - (\imath ql)^N \gamma_N \begin{bmatrix}0 &   \cos N \theta_q  &   \sin N \theta_q   & 0 \\   \cos N \theta_q   & 0 & 0 & 0 \\  \sin N \theta_q   & 0 & 0 & 0\\  0 & 0 & 0 & 0 \end{bmatrix}
\\ \nonumber 
 \gamma_N &=& \chi \zeta S \times \{  (1 +  C^2)  / 4,    C^2/2,  (3/16) (2 C^2 - S^2) \} 
\nonumber \\
 & +  &\chi\zeta    / (1+4 \zeta^2)^N \;  S (1+C^2 )  
 \nonumber \\
 &\;& \; \times \{-1/4,-1/2, -(9/16)(1-4\zeta^2/3)\} \label{gammacoefficients}   
 \end{eqnarray}
  This coupling implies that an initial population of unpolarized particles, in the presence of a spin-orbit (TE-TM) splitting, will after diffusion produce a linearly polarized $S_x, \, S_y$ signal  with a conversion efficiency controlled by $\chi \zeta= l_p / \pi l_{prcsn}$.  This process does not occur without a spin-orbit coupling.   

Production of linear $S_x, S_y$ polarization from an unpolarized source depends very sensitively on whether the system hosts electrons, polaritons, or holes.   First of all,  the polarization has a special angular pattern: a dipole pattern in an electron system, a quadrupole pattern in a polariton system, and a sextupole pattern in a hole-doped system.   Secondly, the signal's magnitude is  strongly sensitive to the spin-orbit interaction, scaling with $ql$ to the $N-th$ power.  At the length scales of interest, much larger than the scattering length $l$, this implies that the linear polarization is strongest for   electrons and weakest for holes.   Thirdly, in the DP regime spin production from an unpolarized source is much smaller for electrons than for polaritons or holes. For electrons the two terms in equation \ref{gammacoefficients}  nearly cancel, giving a signal that scales as $\chi \zeta^3$.  In contrast, for polaritons and holes there is no such cancellation and the signal is proportional to $\chi \zeta$, i.e. larger by two factors of the inverse  dimensionless spin splitting $\zeta^{-1}$, which is large in the DP regime.

Lastly, the linear polarization's dependence on the Zeeman term is  different for each of the three particles.  The difference between particles is  particularly strong in the EY regime where the second term in $\gamma_N$ can be neglected, and relatively weak in the DP regime. We focus on the EY regime in the following discussion.  In electrons the Zeeman term's influence is weak, changing the polarization by at most a factor of $2$.  In contrast, in polaritons removing the Zeeman term (setting $C=0$) extinguishes the linear polarization.    This means that polaritons, unlike electrons and holes, require a magnetic field to produce a linear spin polarization signal from an unpolarized pump. In holes the requirement  is again absent, but one can tune the magnetic field to zero the linear polarization at the resonant condition $|b_z| = |\Delta|/  \sqrt{2}$. As one moves from the EY regime to the DP regime the resonance shifts to smaller $|b_z|$ and eventually disappears. This resonance may be useful for measuring the spin-orbit strength in  hole-doped systems.

Finally we turn to the coupling between linear $S_x, \, S_y$ polarization and circular $S_z$ polarization, $\kappa_{lc}$, which describes  out-of-plane (pseudo)spin precession.   Unlike the couplings $d_{0z}, \kappa_{0l}$ to the number or charge,  the form of $\kappa_{lc}$ depends strongly  on the dimensionless energy splitting $\zeta$, and is independent of the disorder strength $\chi$. It changes if one goes from the DP regime to the EY regime:
\begin{widetext} 
\begin{eqnarray}\label{DiffusonD}
\kappa_{lc} &=& - (\imath ql)^N \begin{bmatrix}0 &  0    & 0   & 0 \\ 0     & 0 & 0 & f_N \cos N \theta_q - g_N \sin N \theta_q \\ 0   & 0 & 0 & f_N \sin N \theta_q+ g_N \cos N \theta_q\\   0 & f_N \cos N \theta_q+ g_N \sin N \theta_q & f_N \sin N \theta_q- g_N \cos N \theta_q & 0 \end{bmatrix}
\nonumber \\
DP: &\,& f_N =0, \; g_N = \zeta S \times \{2, -3, 1 \}
\nonumber \\
EY: &\,& f_N= -CS \, (-2)^{-N}, \; g_N= 0
\end{eqnarray}
\end{widetext}
This coupling implies that a spin-orbit coupling will cause out-of-plane precession, which converts an initially linearly polarized $S_x, \, S_y$ signal to  circular $S_z$ (pseudo)spin, and vice versa.  In the DP regime only a spin-orbit coupling is required, while in the EY regime a magnetic field is also required.     In the DP regime this is bona fide spin precession, since the $g_N$ contribution is anti-Hermitian, similarly to the precession term coupling $S_x$ and $S_y$. In this regime oscillations between $S_x,\;S_y$ and $S_z$ are possible.  In the EY regime oscillations are not possible because  the $f_N$ term is dominant and is Hermitian; we find only a non-oscillatory coupling. 

 Similarly to $\kappa_{0l}$ which produces linear polarization from an unpolarized pump, this precession between linear and circular polarization is strongest for electrons and weakest for holes, and the resulting polarization shows a dipole, quadrupole, or sextupole pattern for electrons, polaritons, and holes respectively.  An interesting difference is that the angular pattern  rotates by $\pi/(2N)$ during the transition from small $\zeta$ to large $\zeta$.   This angular  pattern is expressed in the coupling's dependence on $\sin N \theta_q$ and $\cos N \theta_q$.   These sines and cosines are interchanged when $\zeta$ is increased from the DP limit to the EY limit, causing the rotation.

\subsection{Instability and Regularization}
Our calculation of the response function involved a Taylor series expansion in powers of $ql$.  This is a  long-wavelength  approximation, restricted to length scales larger than the scattering length.  Outside of this approximation's range of validity, i.e.  at momenta $q l > 1$ large enough to probe scales smaller than the scattering length, our analytical formulas predict that the $S_x, S_y$ linear polarization grows exponentially with time instead of decaying, which is an unphysical result.  In polariton systems this instability occurs only  when $\zeta \propto 1/2$ is near the transition between the EY and DP regimes, i.e. when the spin precession length matches the scattering length.

We have regularized this instability by make an ad-hoc modification of the  diffuson, i.e. a cutoff, when $\frac{1}{2}< (ql)^2 $.  This cutoff is technically sound and justified because we already removed all short-distance physics when we truncated higher spatial derivatives.  \cite{PhysRevB.70.155308,PhysRevLett.93.226602}  We have recalculated our numerical results with three different cutoffs, and we compared the results.  When the computed results differ, this signifies that they are sensitive to ballistic physics, i.e. length scales smaller than the scattering length $l$.    Throughout our numerical discussion we report which results are stable, independent of the cutoff, and which results do vary with the cutoff because of a sensitivity to ballistic physics.    Appendix \ref{CutoffAppendix} describes each of the three cutoffs in more detail.

The numerical results presented in our graphs are calculated with a smooth  cutoff that turns on smoothly over the interval $\frac{1}{2} < (ql)^2 < 2$.  This cutoff  nonetheless produces a spatial oscillation, or ringing, in some of our results. The ringing has a length scale set by the cutoff scale $l$ and affects all components of the density matrix.

\subsection{Summary of Analytical Results}
  Remarkably,  electron, polariton, and hole systems  have identical behavior for spatially uniform spin and charge densities. Their spin relaxation rates and precession are exactly the same.  Moreover, there is neither coupling between linear and circular (pseudo)spin, nor between number/charge density and spin.

The behavior of spatially inhomogeneous systems is more intriguing.  An initially unpolarized and non-uniform pump will spontaneously produce both circular and linear polarization.  The resulting circular polarization has the same magnitude for all three types of particles, but for holes and polaritons in the EY regime  it can be zeroed by tuning the magnetic field to a resonance condition where the magnetic field matches the spin-orbit coupling.  The resonance condition differs for holes and polaritons.   In addition to the circular polarization, the linear polarization also can be zeroed in the EY regime by tuning the magnetic field to a resonance with the spin-orbit coupling, but only in holes.  

We have also seen that an initially polarized pump will undergo out-of-plane precession, converting linear polarization to circular polarization and vice versa.  Both this process and the production of linear polarization from an unpolarized pump are strongest for electrons and weakest for holes, and produce distinctive dipole, quadrupole, and sextupole patterns for the three types of particles.  In the case of out-of-plane precession, this pattern's magnitude and angular orientation are both sensitive to the dimensionless energy splitting $\zeta$.
 
The analytical results presented here were obtained with a spatial Fourier transform,  i.e. in momentum space, and they determine the matrix $D^{-1}$ which controls  time evolution.   We now turn to numerical calculation of the response function $\Phi$ in real space.  These numerical results will give us a clear picture of both the spatial distributions that are produced from a spatially localized pump, and of the spin  polarization signals that occur  before and after the spin relaxation time.

\section{\label{sNumerical}Numerical Results}
In the previous analytical section we calculated the diffusion operator $D^{-1}(\vec{q})$ in momentum space, which we presented in equations \ref{DiffusonComponents}-\ref{DiffusonD}.   The response function $\Phi$ in momentum space can not be obtained without numerical calculation, either inverting $\imath \omega + \tau^{-1} D^{-1}$ to obtain the frequency response $\Phi(\omega, \vec{q})$, or exponentiating $\frac{t}{\tau}D^{-1}$ to obtain the time response $\Phi(t, \vec{q})$.   Calculation of the real-space response functions  $\Phi(t, \vec{x}), \; \Phi(\omega, \vec{x})$ requires a further Fourier transform; we discretize on a spatial lattice and use fast Fourier transforms. The lattice spacing $a$ adds an additional  smearing of  the response function $\Phi$ over an area $a^2$, in addition to our diffusive approximation which removes ballistic physics at scales smaller than the scattering length $l$.  If $a < l$ this lattice-induced smearing is insignificant. We employ periodic boundary conditions for numerical convenience, combined with large sample sizes.  In experimental realizations  if  polaritons reach the device edges before they decay, then they escape from the edges of the device.  The behavior of electrons and holes at device edges is sensitive to attached leads.  In the diffusive regime these effects  are not significant for the large sample sizes which we employ here.

\begin{figure}[]
\includegraphics[width=9cm,clip,angle=0]{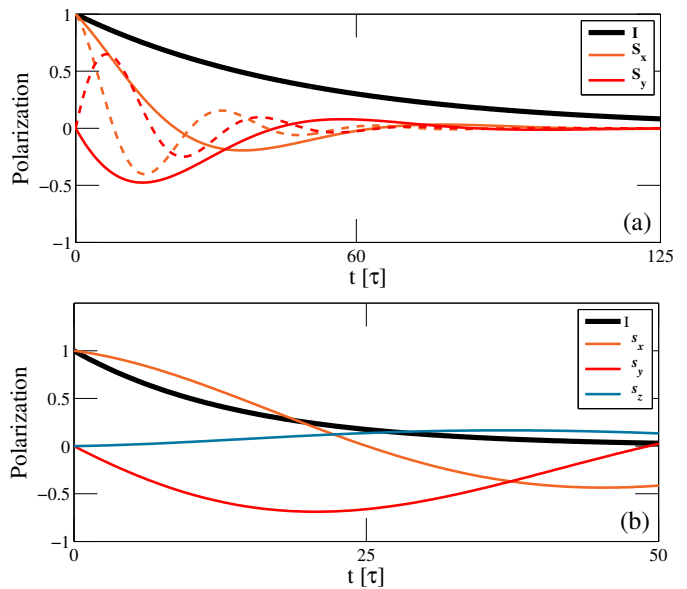} 
\caption{ (Color online.)  Polarization Precession. The  black lines show the polariton density, and the orange, red, and blue lines  show  linear $S_x, \, S_y$ and circular $S_z$ polarization.  Polarization density $S$ is shown in panel a, and polarization degree $s$ is shown in panel b.  The initial polaritons are  $100\%$  linearly $\hat{x}$ polarized.  In panel a the initial polariton density is spatially uniform producing strictly in-plane precession, while in panel b it  has wave-vector $ql=1/\sqrt{10}$ which allows out-of-plane precession. Panel a shows in-plane spin precession - out-of-phase oscillations of the $S_x, \, S_y$ polarizations.   Solid lines show precession with both Zeeman and spin-orbit terms ($\theta_B = 5 \pi /8$), and dashed lines show precession with only a Zeeman term ($\theta_B = 0$.) Panel b shows out-of-plane precession, which produces circular $S_z$ polarization. $\zeta = 0.1, \chi = \hbar / E_p \tau  = 1/4 \pi, \;\tau_{decay} = 50 \tau$, and $\tau = 4 ps$. In panel b $\theta_B = 5 \pi / 8$.}
\label{ProdLifetimeAll}
\end{figure}

\subsection{Basic Spin Precession and Decay}
We begin our numerical results with two graphs that illustrate quantitatively basic features of the diffusion equations in momentum space.  First, Figure \ref{ProdLifetimeAll} studies the spin precession and decay of systems  with fixed wave-vector.   Panel \ref{ProdLifetimeAll}a illustrates precession of an initially $\hat{x}$-polarized density which is  spatially uniform (zero momentum $\vec{q} = 0$) and therefore shows only in-plane and no out-of-plane precession.   The in-plane precession is caused by the Zeeman field oriented perpendicularly to the sample, and is manifested as out of phase oscillations of the  $S_x$ and  $S_y$ linear polarizations (orange and red lines).  Dashed lines show precession with only a Zeeman splitting, and solid lines show precession with both a Zeeman splitting and a spin-orbit term.  The black solid line shows the overall polariton density, which decays exponentially due to the effects of the finite polariton lifetime.  Identical results are obtained for electrons and for holes, which at uniform density  have identical spin relaxation dynamics.  Of course, because the lifetime of electron and hole charge is infinite, their total density is conserved.
   
Panel b of Figure \ref{ProdLifetimeAll} shows out-of-plane spin precession, where  circular $S_z$ polarization is generated from the initial  $\hat{x}$ polarized pump.  Unlike panel a where we plotted the polarization densities $S_x, \,S_y$, here we plot the polarization degrees $s_x, s_y, s_z$, which are  the polarization densities divided by the polariton density $I$.  We will plot  this observable $s$ in all subsequent graphs because it shows out-of-plane precession more clearly than the polarization densities $S$ and moreover can be measured easily in experiments.  In order to obtain out-of-plane precession, the  polariton distribution must have non-zero wave-vector, which we have set to  $\vec{q} = \hat{x} / l \sqrt{10}$.  The largest polarization degrees are found at large times $t \gg \tau$.  At time $t = 50 \tau$ the polariton density is about $3.0\%$ of the initial value,  the initial $100\%$ $\hat{x}$ polarization degree has reversed sign and is $41\%$, and a $13\%$ circular polarization degree has been generated.

\begin{figure}[]
\includegraphics[width=7cm,clip,angle=0]{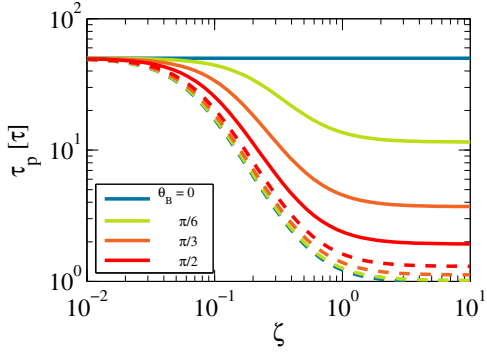}
\caption{ (Color online.)  Polarization (Spin) Decay Times $\tau_p$ as a function of the dimensionless spin splitting $\zeta$ - both for circular $S_z$ polarization (unbroken lines) and for linear $S_x$ polarization (dashed lines.)  We plot $\tau_p$ at four values of $\theta_B = 0,\pi/6, \pi/3, \pi/2$.   In the EY $\zeta \gg 1/2$ regime the decay time is constant, and in the DP $\zeta \ll 1/2$ regime it scales with $\zeta^{-2}$.  $\chi = \hbar / E_p \tau = 1/4 \pi, \;\tau_{decay} = 200 ps = 50 \tau$, and $\tau = 4 ps$.}
\label{ProdLifetimeAllCont}
\end{figure}

In Figure \ref{ProdLifetimeAllCont} we show the polarization decay times  of a spatially uniform distribution  as a function of $\zeta$.   Identical results are obtained for electrons, polaritons, and holes, which in the case of spatially uniform systems have identical spin relaxation dynamics.  We show $\tau_p$ at four values of $\theta_B$.  The D'yakonov-Perel' $\zeta \ll 1/2$ regime is shown on the left side, and  the Elliott-Yafet regime $\zeta \gg 1/2$ regime is shown on the right side. In the EY regime $\tau_p$ is locked to the scattering time $\tau$, and in the DP regime it scales with $\zeta^{-2}$.  At very small $\zeta$  the polariton decay time $\tau_{decay} = 50 \tau$ caps $\tau_p$.  In electron and hole systems this cap is removed and $\tau_p$ continues to scale with $\zeta^{-2}$ even at very small $\zeta$.  
 
 The straight solid blue line in Figure \ref{ProdLifetimeAllCont} shows that when $\theta_B=0$, i.e. when the spin-orbit coupling is zero, the circular $S_z$ polarization lifetime is equal to the polariton lifetime $\tau_{decay}$.  In this case $\sigma_z$ commutes with the Hamiltonian and circular polarization is conserved.  Therefore we find large accumulations of circular $S_z$ polarization when $\theta_B$ is small, i.e. when the spin-orbit coupling is small.

\subsection{The Polarization's Dependence on Radius}

\begin{figure}[]
\includegraphics[width=9cm,clip,angle=0]{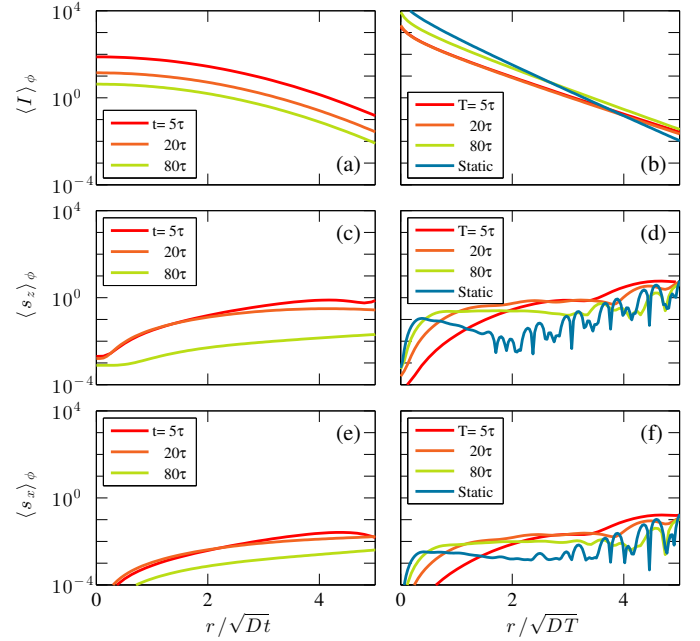} 
\caption{ (Color online.)  Radial Distribution of the Polariton Density and Polarization Degree.  The left panels show the time response ${\rho}(t, \vec{x})$ at time $t$ to a transient pulse, while the right panels show the frequency response to an oscillating signal with period $T$.  All panels show the signal's radial dependence, after performing an rms average over the polar angle $\phi$.  The  radial coordinate is  rescaled by the diffusion length $r_0 =  \sqrt{Dt}$. The upper panels plot the polariton density $\langle I \rangle_{\phi}$ (in arbitrary units) produced by an unpolarized pump, the middle panels plot the circular polarization degree $\langle s_z \rangle_{\phi}$ produced by out-of-plane precession from a linearly polarized pump, and the bottom panels  plot the linear polarization degree $\langle s_x \rangle_{\phi}$ produced by an unpolarized pump.  For polaritons $\tau_{decay} = 50 \tau$.        $\zeta = 0.1, \; \theta_B = 5 \pi / 8, \; \chi = \hbar / E_p \tau  = 1/4 \pi, \;\tau = 4 ps$. }  
\label{ProdRAll}  
\end{figure}

Next we begin our study of the response to a spatially localized pump, which requires numerical evaluation of equations \ref{PhiT} and \ref{PhiOmega}.   In order to learn about typical length scales, Figure \ref{ProdRAll} focuses on  the radial dependence of the polarization.   We present radial data for polaritons only because the electron and hole results are quantitatively similar and qualitatively identical.  The similarity of the radial dependence between the three particles, and indeed all of the qualitative and quantitative results seen in Figure \ref{ProdRAll} and discussed here, was inaccessible without numerical calculations.

 The left panels of Figure  \ref{ProdRAll}  show the temporal response at times $t = 5 \tau, \; 20 \tau, \; 80 \tau$ after the pulse, and the right panels show the frequency response to an oscillating signal with periods $T =  5 \tau, \; 20 \tau, \; 80 \tau$.  The blue lines in the right panels show the response to a static pump.   In the $t,T = 5 \tau, 20 \tau$ calculations we used a system of size $L = W = 51.2 \, l$ and a lattice spacing  $a = 0.2 \, l$.  In the $t,T= 80 \tau$ calculations  $L = W = 102.4 \,l$ and $a = 0.4 \,l$, and in the static calculation $L=W = 204.8 \,l$ and $a = 0.4 \,l$.  Since $a \ll l$, lattice-spacing effects are insignificant. We rms-averaged the intensity $I$ and polarizations $S_x, S_y, S_z$ over  the polar angle $\phi$.  The averaged  polarization degrees $\langle s_x, s_y, s_z \rangle_\phi$ are calculated by dividing the rms-average of $S_x, S_y,S_z$ by the rms-average of $I$.

The upper panels of Figure  \ref{ProdRAll}  plot the polariton density $I$, which is independent of $\zeta$, and also independent of $\theta_B$ and of the starting polarization, except for small variations when $\theta_B = 0$.  In panel a the time-response curves are simple quadratics, and they match perfectly after rescaling $r$ by $\sqrt{Dt}$, which shows that the polariton density is simply an expanding Gaussian.  Similarly in panel b the frequency response curves coincide well after rescaling by $r_0 = \sqrt{DT}$.   The frequency response decays exponentially with $r$ for all $r/r_0 > 0.3$.  (We obtain a good match for the static data by rescaling with the artificial value $T=1280 \, ps$.  This value of $T$ may be tied to the decay time $\tau_{decay} = 200 \, ps$.)   The density's almost perfect insensitivity to $\zeta, \theta_B,$ and time $t,\,T$, after rescaling the radius $r$, supports our use of intensity variables $s_x,\,s_y,\,s_z$ in the remaining graphs.  Because $I$ has a simple profile, dividing by $I$ does not obscure or complicate the physics revealed in the following graphs.

The density's form is very similar to that of  a simple particle density undergoing diffusion without spin-orbit coupling or spin precession.  In this case the diffusion operator is just $D \tau q^2$, where $D$ is the diffusion constant, and  the  Fourier transform to position space can be performed analytically.  For the time response $\Phi(t, \vec{x})$ (Equation \ref{PhiT})  one finds an expanding Gaussian $(2 \pi D t)^{-1} \exp(-r^2 / 2 D t)$.  For the frequency response $\Phi(\omega, \vec{x})$ (Equation \ref{PhiOmega}) one finds a modified Bessel function $2 \pi |K_0(r  \sqrt{\imath \omega /D})|$,  which decays exponentially with $r$ if $ 0.2 < r \sqrt{\omega / D}$.   Our numerical results are very similar to these analytical forms.

The middle panels  c and d  show  out-of-plane precession from a linearly $\hat{x}$ polarized pump to circular $s_z$ polarization degree.   The static response (blue line) shown in panel d displays spatial oscillations which are caused by the regularization that we imposed on the diffusion operator at large wave-vectors $ql \geq 1/\sqrt{2}$.  However we have checked that the overall shape and magnitude of the static response is insensitive to the cutoff. In addition, our calculations using other cutoffs show that the frequency response at $T = 5\tau, \, 20\tau, \, 80 \tau$ is sensitive to the cutoff at small radii $r/\sqrt{DT} < 2$, and other cutoffs not shown here can show substantially larger frequency responses at small $r$.  Outside of this window the frequency response is insensitive to the cutoff.  The temporal response is insensitive to the cutoff everywhere.

Panel c shows that at $t  = 5 \tau, \, 20 \tau$ conversion rates can reach and exceed $10^{-1}$.  Panel d shows that similar numbers are obtained from the response to oscillating and static pumps.   These numbers decrease substantially at larger times or periods $t,T = 80 \tau$, and can be tuned by varying $\zeta$ and $\theta_B$.  An increased polarization degree can also be obtained by moving to larger radii $r$, at the cost of reducing the polariton density. 

Panels e and f show the  linear $s_x$ polarization degree produced in response to an unpolarized pump.  These curves are very similar to the data in panels c and d, but numerically they are about an order of magnitude smaller.  This difference in magnitude  is caused by the factor of $\chi$, the ratio of the energy splitting to the pump energy $E_p$, which controls angular variations in the momentum $k_F$.  $\chi$ is $1/4 \pi \approx 0.08 $ in our calculations.   $\chi$ occurs once in the coupling $\kappa_{0l}$ of polariton density to circular polarization, signalling that this process requires appreciable angular variations of $k_F$.  $\chi$ does not occur at all in the coupling of linear to circular polarization $\kappa_{lc}$; therefore panels e and f are smaller by a factor of $\chi$.

 Our time response and frequency response results  show that the polarization degree - the ratio of the polarization signal to the density - generally increases with $r/r_0$.   This is a universal trend visible in all of our data, for electrons, polaritons, and holes, and for all values of the model parameters, and is seen in panes c,d,e, and f. To be clear, the polarization signal itself decreases with $r$.  However its rate of decrease is smaller than that of the number/charge density $N$, which results in the observed increase in $r$ seen throughout Figure  \ref{ProdRAll}.   At large enough $r$ the polarization degree invariably exceeds $1$, as seen for example in pane d.  However this unphysical result occurs only when the signal intensity is  smaller than the peak value by a factor of $10^{-4}$ or less; it is not physically measurable. This unphysical result is a mathematical artifact of our diffusive approximation, which does not correctly capture the tails of  probability distributions.  In the remaining graphs we restrict $r$ to values  where the diffusive approximation remains valid.

\subsection{Angular structure in position space.}

\begin{figure}[]
\includegraphics[width=9cm,clip,angle=0]{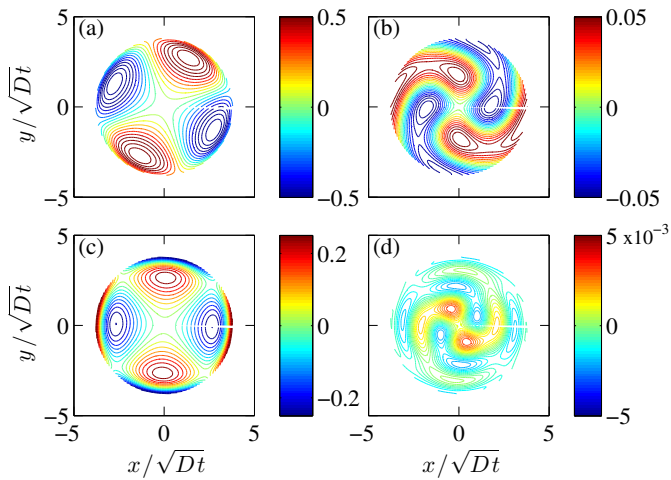} \caption{ (Color online.)  $S_z$ circular polarization degree in real space, at two values of $\zeta$ and two values of time $t$. The left panels and right panels are  $t =  5 \tau$ and $t = 20 \tau$ respectively, and the upper and lower panels are at $\zeta = 0.25$ and $\zeta = 0.5$ respectively.    The pump is  linearly $\hat{x}$ polarized. As $\zeta$ is increased the polarization's quadrupole distribution rotates.   As time increases the quadrupole pattern rotates and deforms into a spiral. $\theta_B = 5 \pi / 8,\; \chi = \hbar / E_p \tau = 1/4 \pi, \;\tau_{decay} = 50 \tau$, and $\tau = 4 ps$. }  
\label{ProdPolarAll}
\end{figure}

We turn to the polarization's angular distribution in real space, which naturally complements the previous section's analysis of the radial dependence.   Our numerical results will confirm the analytical results about the multipole patterns of electrons, polaritons, and holes, and their rotation with $\zeta$.   Furthermore, we will show that the multipoles twist into spirals as time progresses, and we will give quantitative results about the signal strengths. 

\begin{figure}[]
\includegraphics[width=9 cm,clip,angle=0]{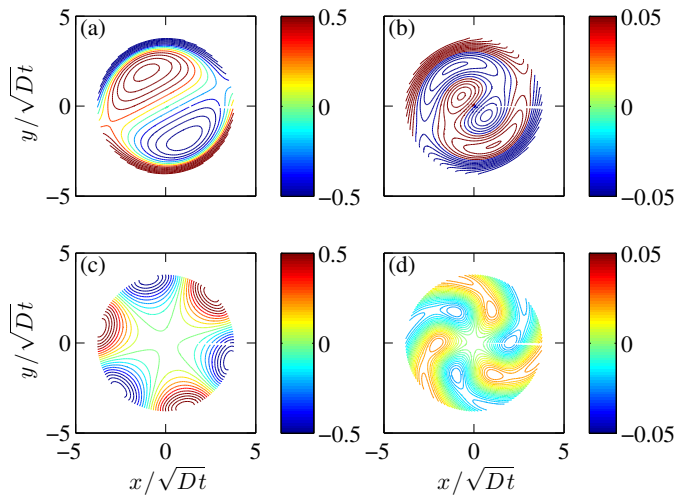} 
\caption{ (Color online.)  $S_z$ spin polarization degree in real space, for electrons (upper panels) and holes (lower panels). The left panels and right panels are  $t =  5 \tau$ and $t = 20 \tau$ respectively.    The pump is linearly $\hat{x}$ polarized.  Electrons show a dipole pattern, and holes show a sextupole pattern.  As time increases the pattern rotates and deforms into a spiral in response to the magnetic field. $\zeta = 0.25, \theta_B = 5 \pi / 8,\; \chi = \hbar / E_p \tau = 1/4 \pi, \tau_{decay}=\infty$, and $\tau = 4 ps$.  }  
\label{ProdPolarCompare}
\end{figure}

Figure \ref{ProdPolarAll} shows the pattern of circular  polarization degree $s_z = S_z/I$  which  out-of-plane precession produces starting from a linearly polarized pump.  Our analytical results indicate that direct conversion of linear to circular polarization in the EY regime requires both a  spin-orbit term and a Zeeman splitting, so we make both terms strong with $\theta_B = 5 \pi/8$. Here and in the other radial plots (Figures \ref{ProdPolarCompare} and \ref{ProdPolarAllCutoff}) the lattice spacing is $0.2 \,l$ and the system size is $51.2 \, l$. Previous theoretical and experimental articles  in the ballistic regime    have shown that out-of-plane precession produces a  quadrupole pattern with two opposite lobes of positive circular polarization, two negative lobes, and the Langbein cross of zero polarization dividing these lobes.    \cite{PhysRevLett.95.136601, Leyder07, PhysRevB.75.075323, PhysRevB.79.125314, PhysRevB.80.165325, PhysRevLett.109.036404, PhysRevB.88.035311}   Further work, always in the ballistic regime, showed that  when a magnetic field is introduced the quadrupole pattern twists into a spiral pattern, with its orientation becoming a function of radius $r$.   \cite{PhysRevB.88.035311, PhysRevLett.110.246403}  This rotation occurs because in the ballistic regime the radial position $r$ is proportional to time $t$, and the external Zeeman field causes in-plane precession at a fixed frequency, which in ballistic fight maps to a precession length.  The combination of the quadrupole pattern with spin in-plane precession causes the spiral. 

As we discussed earlier, our analytical results for the  linear-circular coupling $\kappa_{lc}$ do predict a quadrupole pattern, and in addition indicate that the quadrupole pattern rotates by $\pi/4$ as the $\zeta$ parameter is shifted from the DP regime to the EY regime.
The left panels of Figure \ref{ProdPolarAll} show the circular polarization degree at time $t = 5 \tau$, at $\zeta = 0.25$ (upper left) and at $\zeta = 0.5$ (lower left.)   These panels show a clear quadrupole pattern, and confirm that the pattern rotates as $\zeta$ is changed.  Similarly to the ballistic regime, the rotation is caused by  in-plane precession superimposed on the multilobe pattern.  However - this is a hallmark which can be used to distinguish the diffusive regime from the ballistic regime - no concentric circles are observed in the diffusive regime. Surprisingly, even though there is a Zeeman term the polarization pattern is not twisted into a spiral, unlike previous results in the ballistic regime at $t < \tau$.  It seems that the spiral effect does not occur at this time scale.   However the right panels of Figure \ref{ProdPolarAll} at $t = 20 \tau$ do show clear spiral patterns   at both $\zeta = 0.25$ (upper right) and at $\zeta = 0.5$ (lower right.)  Appendix \ref{CutoffAppendix} shows that these patterns are mildly sensitive to the cutoff but the main features are cutoff-insensitive.  Panel d with its very small polarization degree is the exception: its pattern does depend on the cutoff.   In summary, the circular polarization pattern deforms from a simple quadrupole pattern to a spiral as time progresses, and its angular orientation is sensitive to both $\zeta$ and time $t$.  The most surprising aspect of these results is that no spiral pattern is seen at  time $t = 5 \tau$, unlike ballistic results.  Experimental observation of the  spiral pattern's temporal evolution from a simple multipole to a spiral may be a useful means of measuring the scattering time in individual devices.

 Figure \ref{ProdPolarCompare} shows the angular patterns produced by electrons (top panels) and holes (lower panels), which have been little discussed in the previous literature.  As expected, electrons display a dipole pattern in real space, and holes show a sextupole pattern.  Comparison of the left $t = 5 \tau$ and right $t = 20 \tau$ panels shows again a time dependence - the initial patterns twist into spirals in response to the magnetic field.  In results not shown here, we also have confirmed that the angular orientation of the electron dipole pattern and the hole sextupole pattern rotate in response to variations in $\zeta$, similarly to the polariton's $\zeta$ dependence.  Probably the most interesting aspect of these results is that the observed real-space spin pattern (dipole, quadrupole, sextupole) is a simple signature that can be used to determine the dominant type of spin-orbit interaction.  Moreover, it is likely that in materials with both linear and cubic spin-orbit terms, a numerical analysis of experimental spatial spin distributions  could distinguish both terms in the spin-orbit interaction.

\subsection{The Polarization's Temporal Dependence}

\begin{figure}[]
\includegraphics[width=9cm,clip,angle=0]{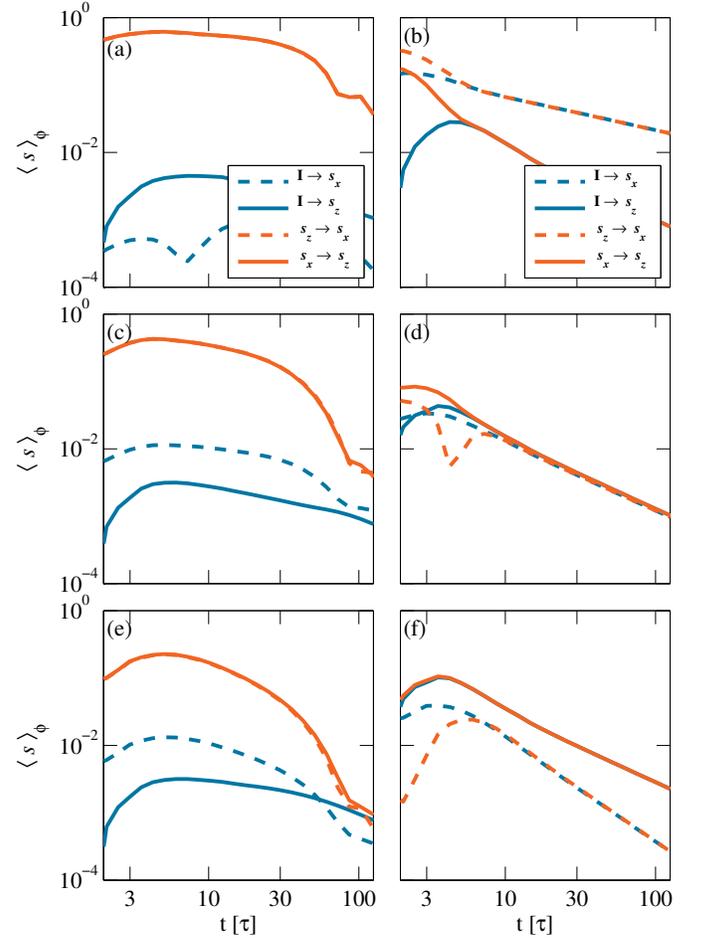}
\caption{(Color online.)  Temporal dependence in electrons (top), polaritons (middle), and holes (bottom). $\zeta = 0.1$ in the left panels and $\zeta = 4$ in the right panels.    Blue lines show the polarization degree produced by an unpolarized pump, and orange lines show  the polarization degree caused by out-of-plane precession from polarized pump.  Dashed lines show linear $s_x$ polarization degree, and solid lines show circular $s_z$ polarization degree.  The data is rms-averaged over the polar angle $\phi$.  For polaritons $\tau_{decay} = 50 \tau$ and for electrons and holes $\tau_{decay} =\infty$. $r/\sqrt{Dt}$ is kept fixed at $3$, $\theta_B = 5 \pi / 8, \; \chi = 1 / E_p \tau = 1/4 \pi, \;\tau = 4 ps$.  }
\label{ProdTimeAll}
\end{figure}

In the previous sections we analyzed the spin's spatial profile; both its radial and angular dependence.  Here we turn to the time dependence of the spin polarization degree, which is determined by the response function $\Phi(t)$, the exponential of the diffusion operator.  In Figure \ref{ProdTimeAll} we examine the time dependence of the spin polarization degree of electrons (top panels), polaritons (middle), and holes (lower panels),  both in the DP regime (left panels) and in the EY regime  (right panels.)   In each panel the dashed and solid lines show respectively the linear $s_x$ polarization degree and the circular $s_z$ polarization degree.  Blue lines show the $s_x, \; s_z$ polarization degrees produced by an unpolarized pump, and orange lines show the polarization degrees caused by out-of-plane precession from a polarized pump.     The lattice spacing $a$, both here and in Figures \ref{ProdXiAll} and \ref{ProdThetaB}, is $a = 0.2 \, r_0 = 0.2 \, \sqrt{Dt}$. Since the lattice spacing is effectively the size of the spatially localized source pump, this means that we are varying the pump size, with maximum width at $t = 125\, \tau$ of about $a \approx 1.6 \; l$. However the effect of this smearing remains small in our plots, which show the spin polarization degree at much large distances $r = 3 \, r_0$.   The sample width and length are $12.8 \, r_0$, large enough to make finite size effects quite small.  We set $\theta_B = 5 \pi / 8$, i.e. both the Zeeman splitting and the spin-orbit splitting are significant, but not precisely in balance.  This allows both in-plane and out-of-plane precession to occur, and therefore gives larger signal strengths than those seen at, for instance, $\theta_B=0$ or $\theta_B = \pi/2$.  At times up  $t = 4 \tau$  the data may be influenced by  ballistic effects which we omitted in our diffusive approximation, and in some cases is sensitive to the cutoff.  In particular, with other cutoffs some spin signals are increased. These ballistic/cutoff effects at small times have little effect on the data after $t = 4 \tau$.

The spin signals in Figure \ref{ProdTimeAll} generally peak at $t =4-8 \tau$ in the DP regime and at $t=1.75-6\tau$ in the EY regime.        Obviously this time scale is determined by the scattering time $\tau$.    At this time scale, in the DP regime, the  out-of-plane precession from an initially polarized pump (orange lines) is dominant, an order of magnitude larger  than polarization generated from an unpolarized pump (blue lines).  In the EY regime the two processes  have similar magnitudes, because any initial polarization decays extremely quickly.    The peak polarization degree is greatest for electrons: $61\%$ in the DP regime, and $32\%$ in the EY regime.  For polaritons it is $43\%$ and $8\%$ respectively for the DP and EY regimes, and for holes it is $23\%$ and $11\%$.    These peak values are independent of the cutoff in the DP regime.  They are more sensitive in the EY regime where the peak occurs earlier.  As a general trend, the polarization signals are invariably larger in the DP regime than in the EY regime. 

Starting at $t =4-8 \tau$ in the DP regime and at $t=1.75-6 \tau$ in the EY regime, the polarization degree decreases uniformly.   At the same time the dashed lines merge with each other, and  solid lines also merge with each other.  This means that the system loses its memory of the initial polarization,  and is  sensitive only to the original polariton intensity.   The memory loss  implies a reversal in the dominant mechanism of spin production: before the memory loss out-of-plane precession is dominant, while after the memory loss it is replaced by  spin generation from an unpolarized source.  

In the EY regime (right panels) the memory loss completes before $t = 10 \tau$, because the spin is tightly locked to the momentum and is randomized at every scattering.  In the DP regime (left panels) the memory loss lasts much longer - $100 \tau$ or longer - because the spin is little affected by scattering and instead is lost by spin-orbit precession around a randomized precession axis.  This is consistent with the analytical expectation that in the DP regime the spin relaxation time scales with $\tau / \xi^2$. In our case with $\xi=0.1$ the memory loss occurs a factor of $10$ later than the time of the peak in the polarization degree.    

Surprisingly, after memory of the initial polarization is completely lost all polarization degree signals do not decay exponentially, but instead as powers of the inverse time (straight lines on our log-log plots).  This power law decay allows the polarization degree to persist to quite long times: at $t=100\tau$  the $s_z$ polarization degree  for polaritons  is  $0.5\%$ in the DP regime, and for electrons it is $7\%$.  In the EY regime the $t=100\tau$  $s_x$ polarization degree  is smaller, but is still $2\%$  in the case of electrons.  

 It is also interesting that after the memory loss the $s_x$ polarization degree is largest for electrons and smallest for holes, because $\kappa_{0l}$'s out-of-plane precession  between $S_x, \, S_y$   and $S_z$   is linear for    electrons, quadratic for polaritons, and cubic for holes.  In particular, in polaritons the $s_x$ and $s_z$ signals are almost the same because in this case all couplings are quadratic in the momentum.    In contrast, the perpendicular $s_z$ polarization degree is  roughly the same for all particles, in keeping with the  $d_{0z}$ coupling which is the same up to a numerical constant.

\subsection{Dependence on the Dimensionless Energy Splitting $\zeta$.}

\begin{figure}[]
\includegraphics[width=7cm,clip,angle=0]{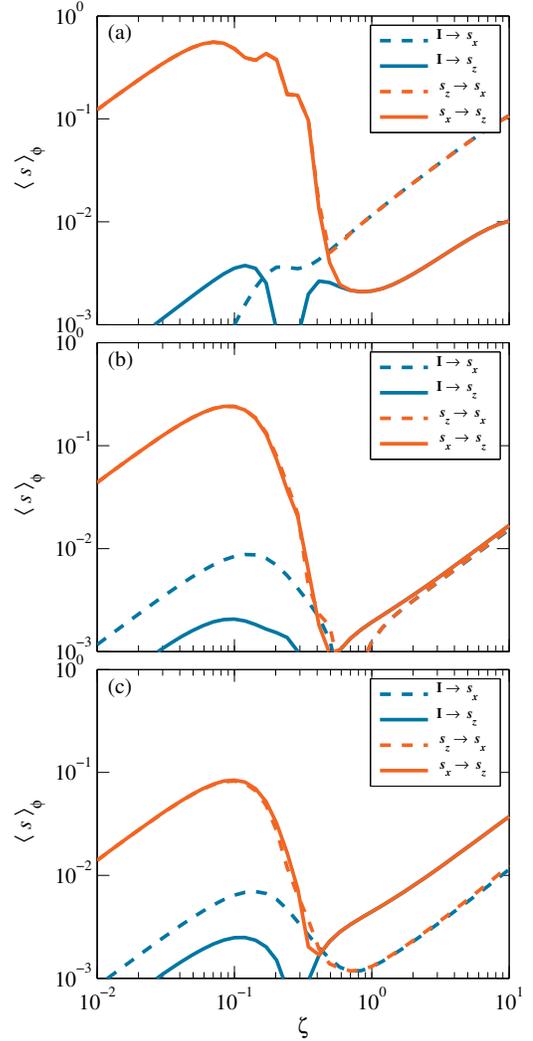}
\caption{ (Color online.)  Dependence on the dimensionless energy splitting $\zeta = l_p / l_{prcsn}$ in electrons (top), polaritons (middle), and holes (bottom).  Similarly to Figure \ref{ProdTimeAll}, blue lines show the polarization degree produced by an unpolarized pump, and orange lines show  the polarization degree caused by out-of-plane precession from a polarized pump.  Dashed lines show linear $s_x$ polarization degree, and solid lines show circular $s_z$ polarization degree.    The data is rms-averaged over the polar angle $\phi$.   Here  $r/\sqrt{Dt}$ is kept fixed at $3$, $t =  20 \tau, \; \theta_B = 5 \pi / 8, \; \chi = 1 / E_p \tau = 1/4 \pi,  \;\tau = 4 ps$.}  
\label{ProdXiAll}
\end{figure}

In Figure \ref{ProdXiAll}  we examine the polarization degree's dependence on the energy splitting $\zeta$, which is the ratio of the scattering length $l$ to the precession length $l_{prcsn}$.  We will confirm the analytical results' linear dependence on $\zeta$ both in the DP and EY regimes.  Going beyond the analytical results, we will reveal the behavior at the transition between DP and EY regimes, give quantitative magnitudes of the signals, and find an interesting feature of polaritons in the EY regime.  Again we have set $\theta_B = 5 \pi / 8$ to  allow both in-plane and out-of-plane precession to occur, and to  give larger signal strengths.  The time is fixed at $t = 20 \tau$, long after the first scattering.  We plot the same observables seen in Figure  \ref{ProdTimeAll}, with the same parameter values.   The data between $\zeta = 0.25$ and $\zeta =1$ is mildly dependent on the cutoff, but the qualitative trends are insensitive to the cutoff and to ballistic physics.

The polarization degree signals show a transition which begins near $\zeta = 0.1$ and ends near $\zeta = 1/2$, i.e. at the transition from the DP regime to the EY regime.  This transition is controlled by the onset of spin relaxation, which occurs at $t \propto \tau / \zeta^2$ in the DP regime, suggesting that  the transition in $\zeta$ scales with $t^{-1/2}$.  At small $\zeta < 0.1$, i.e. deep in the D'yakonov-Perel' regime, the out-of-plane precession signals (orange lines)  are largest and tend to have very similar values.  These signals rise steadily with $\zeta$, in agreement with the linear-circular coupling $\kappa_{lc}$ which is linear in $\zeta$ in the DP regime. Production of $s_x$ linear polarization and $s_z$ circular polarization from an unpolarized pump (the blue  lines) is smaller.  In polaritons and holes this process is linear in $\zeta$, again in agreement with the $\kappa_{0l}$ and $d_{0z}$ couplings which are linear in $\zeta$.  

 Near $\zeta = 0.1$ the out-of-plane precession signals (orange lines) reach peak values of $43\%$ in electrons, $24\%$ in polaritons, and  $8\%$ in holes.   Between $\zeta = 0.1$ and $\zeta = 1/2$ the out-of-plane (orange) signals  drop precipitously, and for polaritons and holes the signals from an unpolarized pump (blue lines) also decrease.   Near the transition to the EY regime at $\zeta = 0.5$ the solid lines merge with each other, as do also the dashed lines.  The mergers indicate that the system    loses its memory of the pump polarization; it is sensitive only to the initial pump intensity, not to the initial polarization.  This is because at $t =  20 \tau$ a system in the EY regime is insensitive to its starting polarization.  In the EY regime all polarization signals are linear in $\zeta$, as expected from our couplings $\kappa_{0l}, \, \kappa_{lc}, \, d_{0z},$ which are all linear in $\zeta$ in the EY regime.  We again see that the $s_x$ polarization degree is largest for electrons and smallest for holes, because $\kappa_{0l}$'s out-of-plane precession  between $S_x, \, S_y$   and $S_z$   is linear for    electrons, quadratic for polaritons, and cubic for holes. In contrast, the perpendicular $s_z$ polarization degree is  roughly the same for electrons, polaritons, and holes.  Interestingly, polaritons in the EY regime give almost exactly equal magnitudes for all four of the signals plotted here.
 

\subsection{ Dependence on $\theta_B$, the balance between  spin-orbit and Zeeman terms.}

\begin{figure}[]
\includegraphics[width=9cm,clip,angle=0]{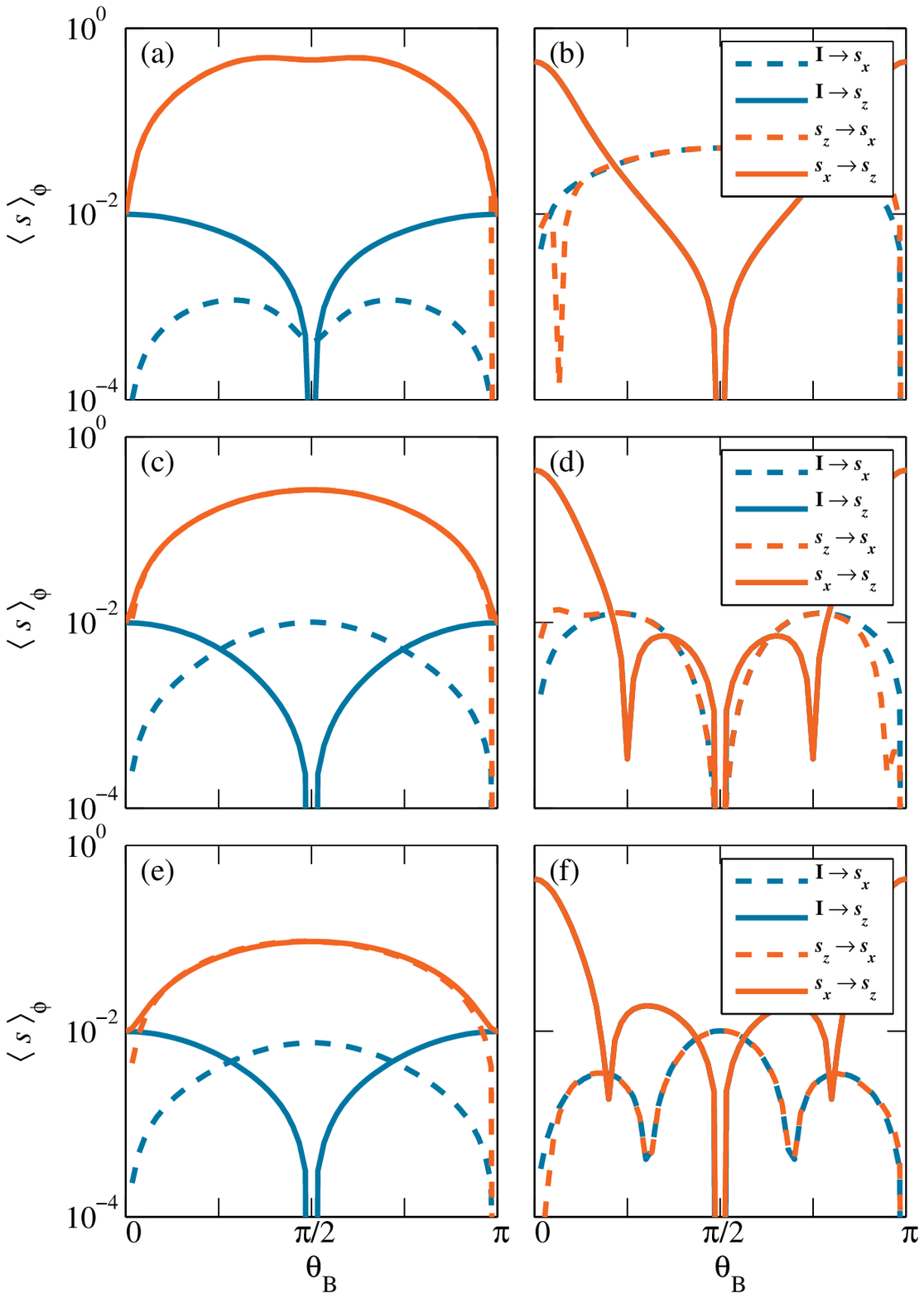}
\caption{ (Color online.)  Dependence on  $\theta_B$ in electrons (top), polaritons (middle), and holes (bottom). $\zeta = 0.1$ in the left panels and $\zeta = 4$ in the right panels.    Blue lines show the polarization degree produced by an unpolarized pump, and orange lines show  the polarization degree caused by out-of-plane precession from a polarized pump.  Dashed lines show linear $s_x$ polarization degree, and solid lines show circular $s_z$ polarization degree.  The data is rms-averaged over the polar angle $\phi$.   $r/\sqrt{Dt}$ is kept fixed at $3$, $t = 80 ps = 20 \tau, \; \chi = \hbar / E_p \tau = 1/4 \pi,  \;\tau = 4 ps$.   }  
\label{ProdThetaB}
\end{figure}

Figure \ref{ProdThetaB}  examines the polarization degree's dependence on the angle $\theta_B$, which describes the balance between the spin-orbit interaction and the Zeeman splitting.  When $\theta_B = 0$ there is only a Zeeman splitting and no spin-orbit interaction, and when $\theta_B = \pi /2$ this is reversed.   This section  confirms the $\theta_B$ dependence which we discussed in the analytical section;  its main value is to illustrate the analytical results and give quantitative magnitudes.  The time is fixed at $t =  20 \tau$, long after the first scattering, and the data is almost completely insensitive to cutoff effects.  We plot the same observables seen in Figures  \ref{ProdTimeAll} and \ref{ProdXiAll}, with the same parameter values.   

If one flips the sign of both the spin-orbit term and the Zeeman splitting, or equivalently adds $\pi$ to $\theta_B$, then the polarization signal may change its overall sign but its magnitude must be unchanged.  Therefore  our plots of the rms-averaged signal are symmetric under $\theta_B \rightarrow \theta_B + \pi$, and we plot only the interval $\theta_B = \left[ 0, \pi \right]$.  In addition we find that three of the four signals in Figure \ref{ProdThetaB} are symmetric under flips around $\theta_B = \pi /2$.  These signals are independent of the sign of the Zeeman term, which distinguishes between clockwise and counterclockwise precession in the device plane.  The one exception occurs in the EY regime, where the linearly polarized $s_x$ response to a circularly polarized $s_z$ pump (dashed orange lines) is not symmetric under flips about $\pi/2$, i.e. it is sensitive to the Zeeman term's sign, as a result of the pump's  orientation parallel to the $z$ axis.  This sensitivity occurs only if the system retains memory  of the original polarization, i.e. either at short times, or even at long times when $\theta_B$ is close to zero as seen here.   However, even this signal is symmetric under changes at the same time  of both the sign of the pump's $s_z$ polarization and also sign of the Zeeman term.

A first examination of Figure \ref{ProdThetaB} shows that the polarization signals are strongly sensitive to $\theta_B$.  In many of our other graphs we have fixed $\theta_B = 5 \pi / 8$ in order to avoid the deep minima that occur at  $\theta_B = 0,  \pi/2,$ and elsewhere.  We also see again that in the EY regime  (right panels)  the solid lines coincide, and the dashed lines also coincide, except near $\theta_B = 0, \pi$ where $S_z$ is conserved.  As noted before, this means that the system has forgotten its   initial state.

As discussed earlier, a special feature of electrons is that in the DP regime the conversion from unpolarized pumps to linear polarization scales with $ \chi \zeta^3$, a factor of $\zeta^2$ smaller than the conversion to circular polarization, and also a factor of $\zeta^2$ smaller than the same signals in polaritons and holes.  In Figure \ref{ProdThetaB}a we see that the unpolarized to linear signal (the dashed blue line) is smaller, but not by the factor of $\zeta^2 = 10^{-2}$ that might be expected.  The reason is that here we are seeing the result of a two step process: (a) conversion of to circular polarization, and (b) precession from circular to linear polarization.  This also explains why the unpolarized to linear signal is zero at $\theta_B = \pi/2$, i.e. when there is no magnetic field.

Broadly speaking, the details of Figure \ref{ProdThetaB}  confirm features of the $\theta_B$ dependence which we found in our analysis of the couplings $\kappa_{0l}, \kappa_{lc}, d_{0z}$.  In particular:
\begin{itemize}
\item No linearly polarized signal is produced when there is no spin-orbit coupling, i.e. at $\theta_B = 0$.  This is universal for electrons, polaritons, and holes, and occurs because a Zeeman field oriented along the $\hat{z}$ axis is unable to create a linear polarization, becuase it is unable  break the rotational symmetry in the $xy$ plane. 
\item  At $\theta_B=0$ the absence of a  spin-orbit term ensures that the  lifetime of circular $S_z$ polarization is infinite. Therefore at $\theta_B = 0$ we find large circular $s_z$ polarization degrees.  The $s_z$ magnitude at $\theta_B = 0$ is close  to $\chi \zeta = \zeta/4 \pi \approx 0.08 \times \zeta$, and is insensitive to $\zeta$ and to any initial linear $S_x$ polarization.  It is also the same for electrons, polaritons, and holes.
\item In the DP regime (left panels), regardless of the value of $\theta_B$, a linearly $\hat{x}$  polarized pulse always generates $s_z$ polarization.  When there is no spin-orbit coupling, i.e. $\theta_B=0$, the $s_z$ polarization is generated by conversion from the pump's initial number/charge density, and its magnitude is of order $\chi\zeta$.  When there is no magnetic field, i.e. $\theta_B = \pi/2$, the spin-orbit coupling causes out-of-plane precession and produces the $s_z$ polarization.  In this case its magnitude reaches $48\%$ in electrons, $27\%$ in polaritons, and $9\%$ in holes. 
\end{itemize}

The most interesting aspects of Figure \ref{ProdThetaB} are the differences between electrons, polaritons, and holes:
\begin{itemize}
\item In the EY regime (right panels) polaritons exhibit no polarization signal at all unless there is a Zeeman term, so at $\theta_B = \pi/2$ all signals are zero. This contrasts with electrons and holes in the EY regime, where a Zeeman term is not necessary for producing an $s_x$ signal, and is required only for producing $s_z$.  This means that for electrons and holes the dashed lines are nonzero at $\theta_B = \pi/2$.
\item In both the EY and the DP regime an unpolarized pump does not produce a linearly polarized $s_x$ signal  unless a magnetic field is introduced, so at $\theta_B =  \pi/2$   the solid blue lines are zero.  
\item Setting $\theta_B = \pi/4,\; 3 \pi/4$ creates a resonance condition where  the Zeeman and spin-orbit terms have precisely the same magnitude, i.e. in polaritons $|b_z| = |\Delta k_F^2|$.  For polaritons  in the EY regime this resonance condition zeroes the generation of  circular $s_z$ polarization f (graphed as  solid blue and orange  lines), causing sharp dips   at $\theta_B = \pi/4,\; 3 \pi/4$.      Holes in the EY regime also display a resonance effect, with a shifted resonance  which occurs at   $|b_z| = \sqrt{2} |\Delta k_F^3|$. In addition, holes  manifest a similar resonance condition in the generation of $s_x$ linear polarization  from an unpolarized  pump, shown as  dashed blue and orange lines.     In contrast electrons do not display any resonance at any value of $b_z$.   The resonances shown here may be useful for measurements of the spin-orbit strength in polariton and hole systems in the EY regime.  
\end{itemize}

\subsection{Summary of the Numerics}
The main value of our numerical results, as opposed to our analytical formulae, is that they offer a clear picture of the spatial spin patterns  in these systems, and also of their time evolution.   Our spatial data on the number/charge density showed that the frequency response always follows a simple decaying exponential in real space, while the temporal response follows a Gaussian.  Moreover, our data showed that changing the time parameter, or indeed any other parameter, has no influence on the number/charge density other than rescaling its exponential or Gaussian profile.

The spin polarization shows much richer behavior.  Electrons, polaritons, and holes produce dipole, quadrupole, and sextupole patterns respectively, which provides a very simple and strong diagnostic tool for visually determining the dominant spin-orbit term.  Our numerical results confirm that these patterns are clearly visible in the diffusive regime.   In real materials the spin-orbit interaction may involve more than one term, for instance both linear and cubic terms, and in this case  numerical analysis of experimental data will reveal the relative strength of the spin-orbit terms.   Our numerical results also showed that in the presence of a magnetic field,  the spin distribution   evolves with time from a simple multipole  into a spiral.  This too will be useful for measuring the scattering time in individual devices.  Moreover, we saw that in the diffusive regime there is no pattern of concentric circles around the pump, which allows an unambiguous determination of whether the system is in the diffusive or ballistic regime.

The numerical data also gave us rich information about the time dependence, and told us how large the  polarization degree can become.  We can break the time evolution into two phases: before and after the time when the system loses memory of its initial polarization.  This memory loss occurs at $t \propto \tau / \xi^2$ in the DP regime, and before $t=10 \,\tau$ in the EY regime.   Before memory loss occurs, the dominant source of spin dynamics is out-of-plane precession which converts $S_x, S_y$ spin to $S_z$ spin, and vice versa.   With an  optimal value of the dimensionless spin splitting $\xi=0.1$, this process yields maximum polarization degrees of $23-61\%$ in the the DP regime at around $t=4-8 \,\tau$.   In the EY regime at $\xi=4$ the maximum polarization degrees  are between $8$ and $32\%$, occur at $t=1.75-6\, \tau$, and  grow linearly with  $\xi$. These maximum values are  largest for electrons.  After the maximum, spin relaxation causes a steady decrease in the polarization degree.

As the system loses its memory of the initial polarization, the dominant mechanism producing  spin polarization changes from spin precession  to direct conversion from the number/charge density.    In this phase the spin polarization degree decreases as a power of the time, and in particular the linear $s_x$ polarization degree is largest for electrons and weakest for holes.  The circular polarization degree is about the same for all three particles.  Even at $t = 100 \tau$ the polarization degree for electrons can remain as high as $7\%$ in the DP regime, or $2\%$ in the EY regime.

Lastly, our numerical results about the $\zeta$ dependence gave the transition between DP and EY regimes (not shown in the analytical results), and gave an optimal value of $\zeta$.  Our $\theta_B$ results largely illustrated and confirmed the analytical results on this parameter.

\section{\label{sDiscussion}Discussion and Conclusions}
After proper rescaling of time and distance,  in the diffusive regime, spatially homogeneous distributions of electrons, polaritons, and holes act identically.   They all have the same spin relaxation times and  the same spin precession rates, and in all cases their number charge/density  shows the same spatial profile spreading away from the pump.  This is perhaps disappointing, since one might wish to see at this level some clear difference between the bosonic polariton and the fermionic electron and hole, similarly to the difference between Bose-Einstein condensation and fermionic repulsion that is seen at large particle densities.  At small particle densities, in the non-interacting limit, the perturbative treatment utilized here shows no such signal.   There is still some possibility that a Berry phase or other quantum effect could distinguish between the three particles, but there is no such effect even in the weak localization signal,  which is determined by relaxation times that are mathematically identical to the spin relaxation times calculated here.  Any such quantum effect will be rather subtle.

However, using a spatially localized pump will immediately reveal very large differences between the  three particles.  The most important difference is their multipole spatial pattern, which has two lobes for electrons, four for polaritons, and six for holes.  This spatial distribution provides an unambiguous signal to experimentalists for determining immediately which spin-orbit term is dominant, and also provides the raw data for quantitative analysis of the strengths of competing spin-orbit terms.   It is a very strong signal: if the pump is initially linearly polarized, then in the DP regime  a multipole pattern in the circular polarization can be observed, with a polarization degree of  up to $23-61\%$, and can persist for long times.   In the EY regime the same process can produce between $8$ and $32\%$ polarization degree.   Production of linear polarization from an unpolarized pump also produces appreciable multipole patterns.

The three particles also differ in their response to a perpendicular magnetic field.  For electrons and holes, in the EY regime an unpolarized pump always produces a multipole pattern of linear polarization, regardless of the magnetic field strength. In contrast for polaritons in the EY regime no linear polarization is generated from an unpolarized pump unless there is a magnetic field. Moreover  in the EY regime polaritons and holes manifest   a resonance, where no circular $S_z$ polarization is generated from an unpolarized pump, at a specific field strength that matches the spin-orbit strength.  The resonance occurs  at numerically different field strengths for different particles, and varies with spin-orbit strength.    This will allow experimentalists both to measure the spin-orbit strength and to confirm again whether the particle is an electron, polariton, or hole. 

The last major difference between the three particles is in the magnitude of multipole pattern that is produced.  In electrons, polaritons, and holes this scales respectively with the first, second, and third power of the scattering length.  In the diffusive regime studied here the scattering length is small compared to the experimental length scale, so the multipole is strongest for electrons and weakest for holes. One interesting way of testing this trend is to compare, at long times, the linear and circular polarization produced from an unpolarized pump.  The circular signal is rotationally symmetric and its magnitude is about the same for all three particles.  In the case of electrons the circular signal should be smaller than the linearly polarized signal, in the case of holes this  is reversed, and in the case of polaritons the two signals  have very similar magnitudes.

\begin{acknowledgments}
This work was supported by FP7 ITM "NOTEDEV", FP7 IRSES projects "QOCaN" and "POLATER" and by Rannis project "Bose, Fermi and hybrid systems for Spintronics."  A. Pervishko thanks the University of Iceland for hospitality.  We thank T. Liew and S. Morina for insightful discussions.
\end{acknowledgments}

\appendix

\begin{widetext}

\section{\label{AppendixFullExpressions} The diffuson  in the general case, for any value of $\zeta$, and for linear, quadratic, and cubic spin-orbit terms. }
In this article we report on  spin dynamics for the following Hamiltonian:
\begin{eqnarray}
H &=& \frac{k^2}{2m}  +\ b_z \sigma_z+ \Delta k^N \begin{bmatrix}0& e^{-\imath N\phi_k} \\  e^{\imath N \phi_k} & 0\end{bmatrix} 
=\frac{k^2}{2m}  +\ b_z \sigma_z+ \Delta \begin{bmatrix}0& (k_x-\imath k_y)^N \\  (k_x+\imath k_y)^N & 0\end{bmatrix}
\end{eqnarray}
We consider the cases of linear $N=1$, quadratic $N=2$, and cubic $N=3$ spin-orbit interactions.  We assume that the impurities conserve the (pseudo)-spin quantum number; if this constraint is relaxed then our results on the matrix structure of the diffuson will change substantially.

For the Hamiltonian $H$, we obtain the diffuson  $D^{-1}$:
\begin{eqnarray}
\tau D^{-1}  &=&\tau \partial_t +   \begin{bmatrix}  (ql)^2/2 & 0 & 0 & d_{0z,N} (ql)^2/2 \\ 0 &  \tau/ \tau_{xx} +d_{xx,N} (ql)^2/2 & \tau/ \tau_{xy} + d_{xy,N} (ql)^2/2 & 0 \\ 0 & -\tau/ \tau_{xy} - d_{yx,N} (ql)^2/2 &  \tau/ \tau_{xx} +  d_{yy,N}(ql)^2/2 & 0 \\ d_{0z,N} (ql)^2/2  & 0 & 0 & \tau/\tau_{zz} + d_{zz} (ql)^2/2 \end{bmatrix} 
\nonumber \\
&-&  (\imath ql)^N \begin{bmatrix}0 &  \gamma_N \cos N \theta_q    & \gamma_N \sin N \theta_q   & 0 \\ \gamma_N \cos N \theta_q     & 0 & 0 & f_N \cos N \theta_q - g_N \sin N \theta_q \\ \gamma_N \sin N \theta_q   & 0 & 0 & f_N \sin N \theta_q+ g_N \cos N \theta_q\\   0 & f_N \cos N \theta_q+ g_N \sin N \theta_q & f_N \sin N \theta_q- g_N \cos N \theta_q & 0 \end{bmatrix}
\nonumber \\
\tau/\tau_{xy}  & = &2C \frac{\zeta  }{1+4\zeta^2}, \;
  \tau/\tau_{xx} = 2 (1 + C^2) \frac{\zeta^2 }{1 +4\zeta^2} , \;
\tau/ \tau_{zz}   =  4 S^2 \frac{ \zeta^2} {1 +4 \zeta^2}, \; d_{zz}  =    (C^2+S^2 \frac{1-12\zeta^2}{(1+4\zeta^2)^3})
\nonumber \\
   d_{xy,2}  & = & d_{xy,3} = d_{yx,2} = d_{yx,3} = \frac{2 C \zeta   (-3 +4\zeta^2) }{(1+4\zeta^2)^3},
   \nonumber \\
  d_{xx,2}   &=& d_{yy,2} =  d_{xx,3} = d_{yy,3} =    \frac{1 }{2} (S^2  + (2-S^2) \frac{1-12\zeta^2}{(1+4\zeta^2)^3}),
  \nonumber \\
   d_{xx,1} &=& d_{xx,2} +  e \cos(2 \theta_q),\; d_{yy,1} = d_{yy,2} -  e \cos(2 \theta_q), \; d_{xy,1} = d_{xy,2} + e \, \sin(2 \theta_q), \; d_{yx,1} = d_{yx,2} - e\, \sin(2 \theta_q)
 \nonumber \\
  e &=& \frac{ 2  \zeta^2  S^2 (3 + 6 \zeta^2 + 8 \zeta^4) }{(1 + 4 \zeta^2)^3} 
  \nonumber \\
   d_{0z,N} &=&  -\chi \zeta C^3 + (N-1) \chi \zeta C S^2 - 1     / (1+4 \zeta^2)^2\times \chi  \zeta  C S^2   \,N
  \nonumber \\
 \gamma_N &=& \chi \zeta S \times \{  (1 +  C^2)  / 4,    C^2/2,  (3/16) (2 C^2 - S^2) \} 
 \nonumber \\
 & +  & \chi \zeta S   / (1+4 \zeta^2)^N \times (1+C^2 )   \times \{-1/4,-1/2, -(9/16)(1-4\zeta^2/3)\}    
 \nonumber \\
 f_N &= & \frac{CS \zeta^2}{(1 + 4 \zeta^2)^{N+1}} \times \{ - (6 + 8 \zeta^2), \,  6   + 12 \zeta^2 + 16 \zeta^4 ,   -(5 + 10 \zeta^2 + 32 \zeta^4 + 32 \zeta^6) \}
 \nonumber \\
g_N &=& \frac{ \zeta S}{(1 + 4 \zeta^2)^{N+1}} \times \{2, (-3 + 4 \zeta^2)/2, 1 - 4 \zeta^2 \}
 \end{eqnarray}
The constant contributions to the matrix $\tau/\tau_{xx}, \tau/\tau_{zz}, \tau/\tau_{xy}$, which determine the evolution of a spatially uniform  distribution, are independent of $N$. The diffusion constant $d_{zz}$ is also independent of $N$.

As described earlier, $S = \sin(\theta_B) =  \Delta k_F^N / \sqrt{(\Delta k_F^N)^2 + b_z^2}$ and $C = \cos(\theta_B)= b_z / \sqrt{(\Delta k_F^N)^2 + b_z^2}$ describe the relative strength of the spin-orbit coupling $\Delta k_F^N$ and the Zeeman term $b_z$.  

\section{Numerical Results on the Cutoff Dependence\label{CutoffAppendix}}
\begin{figure}[]
\includegraphics[width=9cm,clip,angle=0]{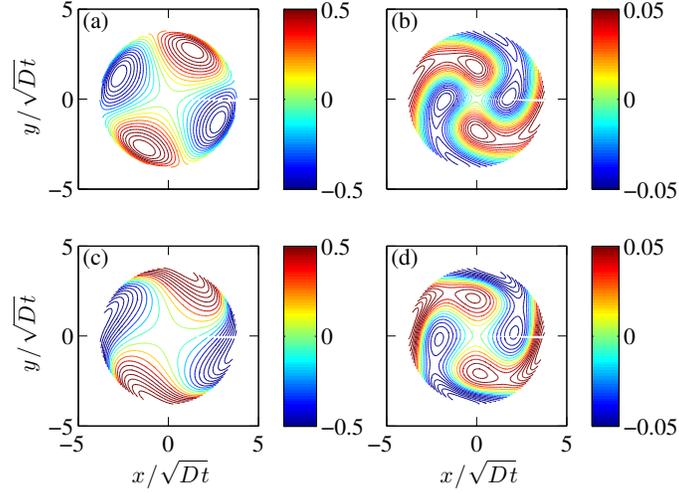} 
\caption{ (Color online.) Cutoff dependence of our results. Circular $S_z$ spin polarization degree in real space, for cutoff A (upper panels) and cutoff B (lower panels). The left panels and right panels are  $t =  5 \tau$ and $t = 20 \tau$ respectively.    The pump is  linearly $\hat{x}$ polarized.   $\zeta = 0.25, \;\theta_B = 5 \pi / 8,\; \chi = \hbar / E_p \tau = 1/4 \pi$, and $\tau = 4 ps$. }  
\label{ProdPolarAllCutoff}
\end{figure}

As discussed in the text, some of our numerical results are sensitive to physics at length scales shorter than the scattering length $l$, i.e. ballistic physics.  These numerical results are changed by our perturbative expansion of the response function in powers of $ql$.  We have tested our results by calculating them in three different ways:
\begin{itemize}
\item Cutoff A: At small wave-number $0.5 > (ql)^2 $ we use the formulas presented in Appendix \ref{AppendixFullExpressions}.  At large wave-number $(ql)^2 > 2$ we set $\kappa_{0l}, \kappa_{lc},$ and $D_{0z}$ to zero, and we set $\Delta=(ql)^2/2$.  At intermediate wave numbers $2 > (ql)^2 > 0.5$ we perform a linear interpolation of  $\kappa_{0l}, \kappa_{lc}, D_{0z}, \Delta$ between these two cases.
\item Cutoff B: At small wave-number $0.5 > (ql)^2 $ we use the formulas presented in Appendix \ref{AppendixFullExpressions}.  At large wave-number $(ql)^2 > 2$ we use the expressions for the Dyakonov-Perel limit, as listed in equations \ref{DiffusonA}, \ref{DiffusonB}, \ref{DiffusonC}, \ref{densitylinear}, and \ref{DiffusonD}.  At intermediate wave numbers $2 > (ql)^2 > 0.5$ we perform a linear interpolation between these two cases. 
\item Cutoff C: We use only the formulas presented in Appendix \ref{AppendixFullExpressions}, without any changes, for all values of $ql$. 
\end{itemize}
Since Cutoff C leaves the analytical formulas unchanged, it retains the original instability near $\zeta=1/2$.  Both Cutoffs A and B do not show instability for electrons and polaritons.   Holes  are stable only with Cutoff A and are unstable with both Cutoffs B and C, presumably because our analytical expressions for holes include cubic $(ql)^3$ terms which are more unstable than the quadratic terms seen in electrons and polaritons.

Figure \ref{ProdPolarAllCutoff} illustrates the differences between two cutoffs. The upper panels are obtained using Cutoff A, and the lower panels with Cutoff B.   Very similar patterns and magnitudes are seen, with some mild distortions.  We have performed similar comparisons, with all three cutoffs, for all of our real-space data.  In cases where  different cutoffs give different results, we have discussed this in the main text.

 \section{\label{AppDiscrepancy} Derivation of the $\kappa_{0l}$ and $d_{0z}$ Couplings} 
Here we report the details of our derivation of the coupling $\kappa_{0l}$ between charge and in-plane $S_x,S_y$ spin, and also in the coupling $d_{0z}$ between charge and out-of-plane $S_z$ spin.  We have performed the calculation systematically in a very general mathematica script which allows calculation to quite high orders in the various small parameters.   However here we do not discuss the script, and instead show in analytical formulas how the derivation develops.  We compute the  $d_{0z}$ coupling which is equal to $-I_{03}$, and the  $\kappa_{0l}$ coupling which is equal   $-I_{01}$ and $-I_{02}$, where $I_{ij}$ is the scattering operator. We begin with equation \ref{JointScatteringOperator}, which writes $I_{ij}$ as the integral
   \begin{eqnarray}
   I_{ij} & = & \frac{\hbar}{4 \pi \nu \tau} \int d\vec{k} \, {Tr}(\, G^A( \vec{k} - \vec{q}/2, E_p)   \sigma_i  
 G^R(  \vec{k} + \vec{q}/2,  E_p ) \, \sigma_j ) 
\end{eqnarray}
 Here $G^A$ and $G^R$  are the disorder-averaged single-particle Green's functions, $\vec{q}$ is the diffuson wave-vector, and $\nu$ is the density of states.  The trace is taken over the spin indices of $G^A, G^R, \sigma_i,$ and $\sigma_j$, which are all $2 \times 2$ matrices in (pseudo)spin space.  The advanced and retarded Green's functions can be resolved into separate contributions from the two  spin states $s=\pm 1$, which have energies $E(s,\vec{k})$:
\begin{eqnarray}
\label{GreensFunctionSpectralRep}
 G^{A,R}(\vec{k}, E) &=& \frac{1}{2}\sum_s \frac{1 + s \vec{X}  (\vec{k} ) \cdot \vec{\sigma} } {E -E(s,\vec{k}) \mp \imath \hbar / 2 \tau},  \; \vec{\sigma} = \left[\sigma_x, \,\sigma_y,\,\sigma_z\right]^T
 \nonumber \\
  X_1 &=& \frac{\Delta\, Re((k_x+\imath k_y)^N)}{ \sqrt{(\Delta |k|^N)^2 + b_z^2}}, \; X_2 = \frac{\Delta\, Im((k_x+\imath k_y)^N)}{ \sqrt{(\Delta |k|^N)^2 + b_z^2}}, \; X_3 = \frac{b_z}{ \sqrt{(\Delta |k|^N)^2 + b_z^2}}
 \end{eqnarray}

 Performing the trace and applying the identity $(ab)^{-1}= (a^{-1}-b^{-1})(b-a)^{-1}$ quickly obtains 
  \begin{eqnarray} \label{DiffusonDetails1}
   d_{0i} & = & -I_{0i} = -\frac{\hbar}{8 \pi \nu \tau} \int d\vec{k} \sum_{s,\acute{s}} 
   (s\,X_i(\vec{k}- \vec{q}/2) + \acute{s}\,X_i(\vec{k}+ \vec{q}/2)) \; Y, 
   \nonumber \\
   Y&=& \frac{1}{E_p -E(s,\vec{k}-\vec{q}/2) - \imath \hbar / 2 \tau} \frac{1}{E_p -E(\acute{s},\vec{k}+\vec{q}/2) + \imath \hbar / 2 \tau}
   \nonumber \\
   & \approx &  
   \frac{\pi \tau}{\hbar}
 \; \left[ \frac{\delta(E_p - E(s, \vec{k}-\vec{q}/2 ) )}{1+ \imath \tau( E(\acute{s}, \vec{k} + \vec{q}/2) - E_p)/\hbar  } + \frac{\; \delta(E_p  - E(\acute{s}, \vec{k}+\vec{q}/2) )}{1 +\imath \tau (E_p -  E(s, \vec{k} - \vec{q}/2))/\hbar  } \right]
\end{eqnarray}
   In the last line we replaced $(E_p -E(s,\vec{k}-\vec{q}/2) - \imath \hbar / 2 \tau)^{-1}$ by its imaginary part, which is taken to be the Dirac delta function $\imath \pi \delta(E_p -E(s,\vec{k}-\vec{q}/2))$.  This standard approximation is justified if the scattering time is long compared to the kinetic time scale, i.e. $\frac{\hbar}{E_p \tau}\propto \chi  \ll 1$.  Physically it means that we concern ourselves only with processes that occur on the elastic scattering circle for polaritons, or on the Fermi surface for electrons and holes.    After these steps  the scattering time  $\tau$ occurs  only in the denominator of $Y$. Powers of  dimensionless energy splitting $\zeta = E_{prcsn} \tau / \hbar$, because they  contain $\tau$, can come only from the denominator of $Y$.  
 
Multiplying $(s\,X_i(\vec{k}- \vec{q}/2) + \acute{s}\,X_i(\vec{k}+ \vec{q}/2))$ by the two terms in $Y$ and making appropriate shifts $\vec{k} \rightarrow  \vec{k} \pm \vec{q}/2$ obtains 
    \begin{eqnarray} \label{DiffusonDetails2}
   d_{0i} & = &- \frac{1}{8 \nu } \int d\vec{k} \sum_{s,\acute{s}}  s  
 \; \left[ \frac{X_i(\vec{k})\;\delta(E_p - E(s, \vec{k} ) )}{1+ \imath \tau( E(\acute{s}, \vec{k} + \vec{q}) - E(s, \vec{k} ))/\hbar  } + \frac{X_i(\vec{k}-\vec{q})\; \delta(E_p  - E(\acute{s}, \vec{k}) )}{1 +\imath \tau (E(\acute{s}, \vec{k}) -  E(s, \vec{k}-\vec{q}))/\hbar  } \right] 
\nonumber \\
   & - & \frac{1}{8 \nu } \int d\vec{k} \sum_{s,\acute{s}}    \acute{s}
 \; \left[ \frac{X_i(\vec{k}+\vec{q})\;\delta(E_p - E(s, \vec{k} ) )}{1+ \imath \tau( E(\acute{s}, \vec{k}+\vec{q} ) - E(s, \vec{k} ) )/\hbar  } + \frac{X_i(\vec{k})\; \delta(E_p  - E(\acute{s}, \vec{k}) )}{1 +\imath \tau (E(\acute{s}, \vec{k}) -  E(s, \vec{k} - \vec{q}))/\hbar  } \right]
\end{eqnarray}
At this point  we expand the $X_i$ functions in powers of $\vec{q}$.  Each power of $\vec{q}$ is implicitly accompanied by a power of $1/k_F$, and after a simple manipulation gives a power of $\chi \, ql$, where $\chi$ is the dimensionless disorder strength and is a small parameter.  Since there are no compensating  $\tau$'s in $X_i$, $\chi$ can be taken at face value as a small parameter, and we expand only to leading order in $\chi$.  In the $d_{0i}$ elements of the diffuson the leading order has a single power of $\chi$, while all other elements of the $D$ are non-zero at zeroth order in $\chi$.  Therefore we write 
\begin{eqnarray}
X_i(\vec{k}+ \vec{q}) = X_i(\vec{k}) + \vec{q}\cdot\vec{\nabla}X_i(\vec{k})
\end{eqnarray}
We decompose the diffuson into two parts, a part $d_{0i}^0$ originating in $X_i(\vec{k})$, and another part $d_{0i}^\zeta$ originating in the gradient $\vec{\nabla}X_i(\vec{k})$:
\begin{eqnarray}
  d_{0i} &=& d_{0i}^0 + d_{0i}^\zeta
 \nonumber \\
  d_{0i}^0 & = &-\frac{1}{8 \nu }\int d\vec{k} \sum_{s,\acute{s}} 
   \;(s+\acute{s}) \left[\frac{X_i(\vec{k})\;\delta(E_p - E(s, \vec{k} ) )}{1+ \imath \tau( E(\acute{s}, \vec{k} + \vec{q}) - E(s, \vec{k} ))/\hbar  } +  \frac{X_i(\vec{k})\; \delta(E_p  - E(\acute{s}, \vec{k}) )}{1 +\imath \tau (E(\acute{s}, \vec{k}) -  E(s, \vec{k} - \vec{q}))/\hbar  } \right]
\nonumber \\
  d_{0i}^\zeta& = &- \frac{1}{8 \nu }\int d\vec{k} \sum_{s,\acute{s}}     
 \; \left[  s \frac{-\vec{q}\cdot\vec{\nabla}X_i(\vec{k})\; \delta(E_p  - E(\acute{s}, \vec{k}) )}{1 +\imath \tau (E(\acute{s}, \vec{k}) -  E(s, \vec{k}-\vec{q}))/\hbar  } + \acute{s} \frac{\vec{q} \cdot \vec{\nabla}X_i(\vec{k})\;\delta(E_p - E(s, \vec{k} ) )}{1+ \imath \tau( E(\acute{s}, \vec{k}+\vec{q} ) - E(s, \vec{k} ) )/\hbar  }  \right]
\end{eqnarray}
This decomposition is physically significant.  The $s \neq \acute{s}$ terms make an exactly null contribution to $d_{0i}^0$, so a little  rearrangement obtains
\begin{eqnarray}
   d_{0i}^0  & =&-\frac{1}{4 \nu }\int d\vec{k} \sum_{s} 
   \;s \left[\frac{X_i(\vec{k})\;\delta(E_p - E(s, \vec{k} ) )}{1+ \imath \tau( E(s, \vec{k} + \vec{q}) - E(s, \vec{k} ))/\hbar  } +  \frac{X_i(\vec{k})\; \delta(E_p  - E(s, \vec{k}) )}{1 +\imath \tau (E(s, \vec{k}) -  E(s, \vec{k} - \vec{q}))/\hbar  } \right]
   \nonumber \\
   & =&-\frac{1}{2 \nu }\int d\vec{k} \sum_{s} 
   \;s \frac{X_i(\vec{k})\;\delta(E_p - E(s, \vec{k} ) )}{1+ \imath \tau \vec{q} \cdot \nabla E(s, \vec{k})/\hbar  } 
\end{eqnarray}
In the last line we have expanded in powers of $\vec{q}$, following the same argument as before.  In contrast to the previous expansion, the first term in the present expansion  is matched by a factor of $\tau$, so it actually scales as $ql$ without any powers of $\chi$.  Since the denominator subtracts $E(s, \vec{k} )$ from $E(s, \vec{k} + \vec{q})$ and therefore has no powers of the spin-orbit splitting,  $d_{0i}^0 $ can contain no powers of $(1 + 4 \zeta^2)^{-1}$.  Because of the guarding factor of $s$ multiplying the sum,  non-zero contributions can be obtained only from a dependence on the spin index $s$ of the Dirac delta function, or of $X_i$, or of $\nabla E(s, \vec{k})$.  Physically, variations in the Dirac delta function with $s$ reflect a difference between the density of states on the spin up $s=+1$ Fermi surface and the density of states on the spin down Fermi surface.  Similarly, variations in $\nabla E(s, \vec{k})$ correspond to differences in the Fermi velocity on the two Fermi surfaces, and variations in $X_i$  to differences in the spin-orbit interaction's angular orientation on the two surfaces.  In order to analyze these effects, it is necessary to calculate the spin-splitting of the two $s=\pm 1$  Fermi surfaces, as a function of the angular variable $\phi_k$.  It is sufficient to obtain them to first order in the ratio of the spin splitting to the Fermi energy, $E_{prcsn}/E_p= \chi \zeta$.  After doing so, we obtain in the case of the  Rashba interaction
\begin{eqnarray}
   d_{01}^0  & =&-\frac{1}{2 \nu }\int d\vec{k} \sum_{s} 
   \;s \frac{X_1(\vec{k})\;\delta(E_p - E(s, \vec{k} ) )}{1+ \imath \tau \vec{q} \cdot \nabla E(s, \vec{k})/\hbar  } 
    \nonumber \\
 & = &   -  \int \frac{ d\phi_k}{2 \nu}\sum_s s \, \rho(s,\phi_k) \, \frac{X_1(s,\phi_k)}{ 1+ \imath \tau \vec{q} \cdot \nabla E(s, \phi_k)/\hbar } 
 \nonumber \\
 \rho(s,\phi) &=& \int k \, {dk} \, \delta(E_p - E(s, \vec{k} ) )= \frac{k_F^2}{2 E_F} ( 1 - \chi \zeta   s \sin^2(\theta_B)/ 2)
 \nonumber \\
 \nu &= & \sum_s \, \int_0^{2\pi} {d \phi}  \rho(s,\phi)  = 4 \pi \frac{k_F^2}{2 E_F}
 \nonumber \\
 X_1(s,\phi) &=&   \cos(\theta_k)\sin(\theta_B)(1 -  s \chi \zeta \cos^2(\theta_B) /2 )
 \nonumber \\
\frac {1} { 1+ \imath \tau \vec{q} \cdot \nabla E(s, \phi_k)/\hbar } &=& 1 -\imath ql  \cos(\theta_k - \theta_q) + \imath ql  \chi \zeta  \cos^2(\theta_B) s \cos(\theta_k - \theta_q)/2
\nonumber \\
  d_{01}^0  & =& - \imath  q_x l \chi \zeta  \sin(\theta_B)(1 +  \cos^2(\theta_B) ) / 4
 \end{eqnarray}
We have  automated this perturbative calculation, and also the  calculation of $d^\zeta$, with mathematica, which facilitates the calculation of all matrix elements to high orders.

We now turn to the other contribution, $ d_{0i}^\zeta$.  Exchanging $s, \acute{s}$ in the second term and the expanding in powers of $\vec{q} \cdot \nabla E \propto ql $ produces
\begin{eqnarray}
  d_{0i}^\zeta& = & -\frac{1}{8 \nu } \int d\vec{k} \sum_{s,\acute{s}} 
   s  \;\vec{q}\cdot\vec{\nabla}X_i(\vec{k})\; \delta(E_p  - E(\acute{s}, \vec{k}) )
 \; \left[  \frac{-1}{1 +\imath \tau (E(\acute{s}, \vec{k}) -  E(s, \vec{k}-\vec{q}))/\hbar  } + \frac{1}{1+ \imath \tau( E(s, \vec{k}+\vec{q} ) - E(\acute{s}, \vec{k} ) )/\hbar  } \right]
\nonumber \\
& = &- \frac{1}{8 \nu }\int d\vec{k} \sum_{s,\acute{s}} 
   s \;\vec{q}\cdot\vec{\nabla}X_i(\vec{k})\; \delta(E_p  - E(\acute{s}, \vec{k}) ) \; 
\sum_M \;\sum_{\pm}  \frac{\mp (-\imath \tau \vec{v}_F(\vec{k}) \cdot \vec{q})^{M-1}}{(1   \pm \imath \tau (E(\acute{s}, \vec{k}) -  E(s, \vec{k}))/\hbar)^{M}  }  
\end{eqnarray}
The $\vec{q}\cdot\vec{\nabla}X_i(\vec{k})$ multiplying the entire expression is proportional to $\chi \, ql$, which guarantees a power of $\chi$.  Therefore all other expressions must remain at zeroth order in $\chi$, allowing us to neglect the spin splitting effects on the density of states, Fermi velocity, and $X_i$, which are so important for determining $ d_{0i}^0$.  In this approximation the Fermi surface becomes a simple circle and the $ d\vec{k} $ turns into a simple angular integral over $\phi_k$.  Moreover, the factor of $s$ multiplying everything guarantees that the two $s = \acute{s}$ contributions sum to zero, leaving us with 
\begin{eqnarray}
  d_{0i}^\zeta& = &- \int \frac{d\phi_k} {16 \pi}   \;
     \vec{q}\cdot\vec{\nabla}X_i(\vec{k})\; \sum_M(-\imath \tau \vec{v}_F(\vec{k}) \cdot \vec{q})^{M-1}
 \;\sum_{\pm} \sum_{s} \frac{\mp s}{(1   \mp \imath s2\zeta)^{M}  }  
\end{eqnarray}
For the couplings to in-plane spins, $  d_{01}^\zeta$ and  $  d_{02}^\zeta$, the angular integration selects the $M=N$ term, where $N=1,2,3$ for electrons, polaritons, and holes.  For the coupling to out-of-plane spins,  $  d_{03}^\zeta$, the angular integral selects the $M=2$ term.
For $N=1,2,3$ the last factor $\sum_{\pm} \sum_{s} \frac{\mp s}{(1   \mp \imath s2\zeta)^{M}  }  $ sums to 
\begin{eqnarray}
\frac{-\imath 8 \zeta}{(1+4 \zeta^2)^N}\times (1,2,3(1-4\zeta^2/3))
\end{eqnarray}
If one performs the remaining angular integral and sums $  d_{0i}^0,   d_{0i}^\zeta$, then one obtains the charge-spin couplings reported in this article.

 \end{widetext}

\bibliography{Vincent}

\end{document}